\documentclass[12pt]{iopart}
\usepackage{amsmath,bm}
\usepackage{amssymb}
\usepackage{graphicx}
\usepackage{amsthm}

\begin{document}

\newcommand \be  {\begin{equation}}
\newcommand \bea {\begin{eqnarray} \nonumber }
\newcommand \ee  {\end{equation}}
\newcommand \eea {\end{eqnarray}}
\renewcommand{\leq}{\leqslant}
\renewcommand{\geq}{\geqslant}
\newtheorem{lemma}{Conjecture}

\newcommand{\Ai}{\operatorname{Ai}}

\newtheorem{proposition}{Proposition}
\newtheorem{conjecture}{Conjecture}

\theoremstyle{definition}

\title[Landscape in the simplest
 random least-square problem]{Optimization landscape in the simplest constrained
 random least-square problem}

\vskip 0.2cm
\author{Yan V. Fyodorov $^{1,2}$ and Rashel Tublin $^1$}
\address{$^1$ King's College London, Department of Mathematics, London  WC2R 2LS, United Kingdom}

\address{$^2$ L.D. Landau Institute for Theoretical Physics, Semenova 1a, 142432 Chernogolovka, Russia}

\maketitle

\begin{abstract}
We analyze statistical features of the ``optimization landscape'' in a random  version of one of the simplest constrained optimization problems of the least-square type: finding the best approximation for the solution of an overcomplete system of $M>N$ linear equations $(\bm a_k,\bm x)=b_k, \, k=1,\ldots,M$ on the $N-$sphere ${\bf x}^2=N$. We treat both the $N-$component vectors $\bm a_k$ and parameters $b_k$ as independent mean zero real Gaussian random variables. First,  we derive the exact expressions for the  mean number of stationary points of the least-square loss function in the framework of the Kac-Rice approach combined with the Random Matrix Theory for Wishart Ensemble, and then perform its asymptotic analysis as $N\to \infty$  at a fixed $\alpha=M/N>1$ in various regimes. In particular, this analysis allows to extract the Large Deviation Function for the density of the smallest Lagrange multiplier $\lambda_{min}$ associated with the problem, and in this way to find its most probable value. This can be further used to predict the asymptotic minimal value ${\cal E}_{min}$ of the loss function as $N\to \infty$. Finally, we develop an alternative approach based on the replica trick to conjecture the form of the Large Deviation function for the density of ${\cal E}_{min}$ at $N\gg 1$.
As a by-product, we find the value of the {\it compatibility threshold} $\alpha_c$ which is the minimal value of the asymptotic ratio $M/N$ such that the random linear system on the $N-$sphere is typically compatible.
\end{abstract}

\section{Introduction}
The aim of this paper is to analyze a random version of one of the simplest, yet non-trivial minimization problems
	 of the least-square type over the sphere ${\bm x}^2=N$, with ${\bm x}\in \mathbb{R}^N$. Namely, given a matrix $A$  with $MN$ real entries (arranged in $M$ rows and $N$ columns), and a vector $\bm b \in \mathbb{R}^M$ we seek for the vector ${\bm x}$ satisfying ${\bm x}^2=N$ which minimizes the following loss/cost function:
	\begin{equation}\label{energy}
		H({\bm x})=\frac{1}{2}||A{\bm x}-{\bm b}||^2:=\frac{1}{2}\sum_{k=1}^M\left[\sum_{j=1}^NA_{kj} x_j-b_k\right]^2
	\end{equation}

Initial motivation to study the above-formulated question came from the area of statistics called \textit{Multiple Factor Analysis} due to Thurston, see \cite{MFAbook}.  One of the paradigmatic problems in  this area is known as the \textbf{Procrustes problem} and is formulated as follows:\\
{\it Given an $M \times N, \, M>N$ matrix $A$ and a ``target structure'' matrix $B$ of the same dimension, one is  asked to find a $N\times N$ matrix $X$ such that $B=AX$ holds with maximal precision, and columns of $X$ are of unit norm}.\\
As is easy to see, the system of equations for entries of $X$ is overcomplete for $M>N$, so in general, one can not find $X$ which satisfies it exactly. To find an approximate solution the best one can do is to minimize some cost function that penalizes deviations from the relation $B=AX$.  Denoting $\bm x$ and $\bm b$ the corresponding columns of $X$ and $B$ respectively, one can seek column by column the best solution as a minimization problem restricted to the unit sphere $\bm x^T  \bm x = 1$. In this setting, the least square fitting is one of the most natural and frequently used minimizations, with the cost function $||A\bm x - \bm b||^2$. As such this problem attracted considerable attention starting from the work \cite{Browne1967}, see e.g. \cite{Gander1981,GolubMatt1991}.

The main innovation of our approach is that we further consider the entries $A_{kj}$ of positive definite $M\times N$ matrix $A$  as independently identically distributed normal real variables such that $A^TA=W$	is $N\times N$ Wishart matrix with the density (with respect to the corresponding Lebesgue measure $dW$)  given by
	\begin{equation}\label{Wishart}
		P_{N,M}(W)=C_{N,M}e^{-\frac{N}{2}\mbox{\small Tr} W}\left( \det W\right)^{\frac{M-N-1}{2}}
	\end{equation}
where above (and in the main part of the paper) we restrict our consideration only to the case $M>N$, relaxing it in the last part of the paper. The components $b_i$ of the noise vector are further assumed to be
	normally distributed: $b_k,\,k=1,\ldots,M$  are i.i.d. mean zero real Gaussian variables with the covariance $\left\langle b_kb_l \right\rangle=\delta_{kl}\sigma^2$, where the notation $\left\langle \ldots \right \rangle $ here and henceforth stands for the expected value $\mathbb{E}[\ldots]$ with respect to all types of random variables. In other words, ${\bm b}=(b_1,\ldots,b_M)^T = {\cal N}({\bm 0},\sigma^2 {\bm 1}_M)$.
Also note that minimizing over a sphere of unit radius can be replaced by any
fixed radius, up to a trivial rescaling. We find it is more convenient to replace the unit norm with the condition  ${\bm x}^2=N$.
 	
In such a setting we aim at addressing the following questions about
 the structure of the``cost/loss function landscape'' associated with eq.(\ref{energy}). In particular, for a given $M,N$ and the noise parameter $\sigma^2$ we would like to
\begin{itemize}
\item[\textbf{i.}] count the stationary points of this function via  the  Lagrange multipliers method.
    \item[\textbf{ii.}] in the limit $(M,N)\to \infty$ with a fixed ratio $\alpha=M/N>1$ analyze statistics of the Lagrange multiplier corresponding to the minimal cost/loss
      ${\cal E}_{min}$, and eventually characterize ${\cal E}_{min}$ by its mean value, variance, and if possible
       the Large Deviation Rate.
\end{itemize}
To achieve those goals we will use two different (and largely complementary) approaches. The first is based on the method of Lagrange multipliers combined with the Random Matrix theory. In that framework we are able to find expected number of different stationary points in the cost function landscape by using the so-called  Kac-Rice formula. Another method exploits ideas from Statistical Mechanics for directly searching and characterizing ${\cal E}_{min}$. When doing this one can dispose of the restriction $\alpha=M/N>1$. Below we give a flavour of these methods and a brief historic account of related studies.

 The associated ''random landscape paradigm'' originated in the theory of disordered systems such as  spin glasses, see \cite{CCrev} for an accessible introduction, and gradually became popular beyond the original setting finding numerous applications in such diverse fields as machine learning via deep neural networks \cite{Choromanska_et_al,Bask1}, and large-size inference problems in statistics \cite{BMMN19,RABC20,MBAB20}. The important information in that case is associated not only with the property of the global minimum of a given ''random landscape'' exemplified by a certain specified function $V(\bm x)$, but also with the number and position of all other stationary points (minima, maxima and saddle points) on the corresponding landscape surface.
 The simplest nontrivial characteristics, the mean counting function of stationary points irrespective of their index, in that setting is given by the so-called Kac-Rice Formula. The counting problem amounts
to finding all solutions of the simultaneous stationarity
conditions $\partial_k  V=0$ for all $k=1,...,N$, with
$\partial_k$ standing for the partial derivative
$\frac{\partial}{\partial x_k}$. The mean total number ${\cal N}_s(D)=\left\langle\#_D \right\rangle$ of the
stationary points in any spatial domain $D\in \mathbb{R}^N$ can  be found according
to the multidimensional integral formula for the number of (isolated) solutions of
a system of $N$ equations in $N$ unknowns $\left\{ f_i(x_1,\dots,x_N)  =0\right\}_{i=1}^N$:
	\begin{equation}\label{KRintro1}
		\#_D = \int_D \delta (f_1)\cdots \delta(f_N) \left| \det\left(\frac{\partial f_i}{\partial x_j}\right) \right| dx_1\dots dx_N,
	\end{equation}
 with $\delta(f)$ standing for the Dirac delta-function, and appropriate smoothness of the functions $f_i(\bm x)$
 is assumed. In our case $f_k(x_1,\dots,x_N)  =\partial_k  V$, so  that the mean of the number of stationary points
can be written as ${\cal N}_s(D)=\int_D \rho_s({\bf x}) \, d{\bf x}$, with $\rho_s({\bf
x})$ being the corresponding mean density of the stationary points given by
\begin{equation}\label{KRintro2}
\left\langle \rho_{s}({\bf x})\right\rangle =\left\langle
|\det\left(\partial^2_{k_1,k_2} V\right)|
\prod_{k=1}^N\delta(\partial_k V)\right\rangle,
\end{equation}
where here and henceforth in the paper the brackets $\langle \ldots \rangle$ stand for taking the
    expectation with respect to all relevant random parameters.
\eqref{KRintro2} is exactly the mentioned Kac-Rice formula, see \cite{Fyo15} for a discussion and further references.

 Note the importance of keeping the modulus of the determinant of the Hessian matrix $\partial^2_{k_1,k_2} V$ in \eqref{KRintro2}, as omitting it would yield instead the density of
the object related to the Euler characteristics of the surface, see the book \cite{math1}.
The presence of such modulus of the Hessian for a long time was considered to be a serious
obstacle preventing general evaluation of the mean number of stationary points. The paper \cite{Fyo04} was seemingly the first relating the counting problem to the Random Matrix Theory (RMT) context. These ideas have been further developed in \cite{BraDea07,FyoWil07,FyoNad12}.  Independently, very similar approach has been rediscovered in \cite{Auf1}, which together with its sequel \cite{Auf2}) considerably advanced that technique and provided important insights into counting stationary points  with a fixed index on a sphere for a broad class of landscapes.
Following that pattern,  the work  \cite{FLD2013}  addressed how the counting (and other landscape properties) change as a function of a certain control parameter for the simplest Gaussian cost landscape on the sphere.

Our present work in a sense adopts the same strategy as \cite{FLD2013}, adjusting it to the case of the least-square landscape.  First, combining the machinery of random matrices with the method of Lagrange multipliers we are able to
\begin{itemize}
\item[i)] rigorously derive the exact (finite $N,M,\sigma^2$)  expression for the  mean number of stationary points of the cost function \eqref{energy} and provide its subsequent asymptotic analysis in various regimes.
 \item[ii)] extract the mean and the large deviation function for the value of the minimal Lagrange multiplier $\lambda_{min}$ of the associated optimization problem as $N\to \infty$ at a fixed ration $\alpha=M/N>1$.
\item[iii)] assuming a certain self-averaging property, compute the (normalized) asymptotic mean of the minimal value $\langle \mathcal E_{min} \rangle$ of the cost function \eqref{energy} finding that it is given by \eqref{minmeanvalue}.
\end{itemize}

    To verify the latter formula for the minimal cost independently (and in this way also to justify some assumptions used to arrive to it) in
 the rest of the paper  we follow \cite{FLD2013} and reformulate the minimization problem from a viewpoint of Statistical Mechanics of disordered systems.
 Namely, to look for the minimum of a loss/cost function $H (\bm x)$ over the sphere ${\bm x }^2=N$ one
introduces an auxilliary non-negative parameter $\beta>0$ (called {\it the inverse temperature}) and uses it to define the so-called {\it partition function} $Z(\beta)$ via
	\begin{equation}\label{method Z_intro}
		Z = \int\limits_{{\bm x }^2=N} d \bm x \, e^{-\beta H(\bm x)}.
	\end{equation}
	Suppose the minimum $\mathcal E_{min}$ of the loss/cost function $H (\bm x)$ is achieved at some $\bm x_{min}$. The crucial point is to  observe that
 applying the Laplace's method for $\beta\to \infty$ gives the leading exponential behaviour in the form
	\begin{equation*}
		Z \propto e^{-\beta \mathcal E_{min}}
	\end{equation*}
hence the minimal value can be recovered in the limit $\beta\to \infty$ from the logarithm of the partition function:
	\begin{equation}\label{min_free_energy_intro}
		 \mathcal E_{min}  = -\lim_{\beta\to\infty}\frac{1}{\beta}  \log Z
	\end{equation}
 This relation is very general (in particular, it does not assume the global minimum is achieved at only a single $\bm x_{min}$, it may be several such points),  but for problems involving random cost/loss functions its actual usefulness crucially depends on our ability to characterize the behaviour of the log in the right-hand side.
 The simplest, yet already highly nontrivial task is to find the mean value  $\langle \mathcal E_{min} \rangle$ for the minimal cost. Obviously, one needs to average the logarithm $ \log Z$ in the right-hand side. That field of research originated in the Physics literature as the ``theory of spin glasses'' \cite{CCrev}.  Doing this fully rigorously even in the simplest instance is quite challenging, though considerable progress has been achieved in the last decades in evaluating such averages in a mathematically controllable way in the case when the cost function is normally distributed,  see e.g.~\cite{TalSpinGlass}.  Unfortunately, in our case the cost function is not normally distributed, but rather represents a sum of squared normally distributed pieces. In such a case the rigorous theory has not been yet developed, but progress is still possible within the powerful but heuristic method of Theoretical Physics, known as the ''replica trick'', see e.g. \cite{PUZ2020}.  This approach assumes  that the mean value we are after can be found not from directly calculating the average $\langle \log Z  \rangle$ but by considering the expectation of the integer moments of the partition function, frequently called in the physical literature the ''replicated'' disorder averaged partition function $\langle Z^n  \rangle$  and subsequently taking the limit $n\to 0$ to recover the averaged log:
	\begin{equation}\label{replica method_intro}
		\langle  \mathcal E_{min} \rangle = \lim_{\beta\to\infty}\frac{1}{\beta}\lim\limits_{n\rightarrow 0} \frac{1}{n} \log{\langle Z^n \rangle}
	\end{equation}
The details of evaluating moments $\langle Z^n  \rangle$ for our problem in a closed-form will be presented in Sec. \ref{Large deviations for the minimal cost} and  the Appendix \ref{partion_fun_evaluation}. In fact, as was observed in \cite{FLD2013}, the replica trick sometimes can be used not only to calculate the mean value of the global minimum but also to characterize fluctuations around it for large $N\gg 1$, employing the Large Deviations approach. Below we briefly give an account of that idea following the above paper.

One starts with assuming that the random variable ${\mathcal E}_{min}$  is characterized by a probability density ${\cal P}_N({\mathcal E}_{min})$ which has for large $N\gg 1$ a Large Deviations form:
\begin{equation}\label{mean_Large_dev}
{\cal P}_N({\mathcal E}_{min})\approx R({\bm e})e^{-N{\cal L}({\bm e})}, \quad {\bm e}={\mathcal E}_{min}/N
\end{equation}
with the rate ${\cal L}({\bm e})$ and a leading prefactor $R({\bm e})$.
On the other hand, consider again the ``replicated'' disorder averaged partition function $\langle Z^n  \rangle$,
but instead of considering the limits $n\to 0$ and $\beta\to \infty$ separately, let us make such a limit
by keeping the product $n\beta=:s$ fixed. In this way we may write
\begin{equation}\label{LD1}
\lim_{n=s/\beta, \beta\to \infty}\langle Z^n  \rangle=\lim_{n=s/\beta, \beta\to \infty}\langle e^{n\log Z}  \rangle=\lim_{\beta\to \infty}\langle  e^{s/\beta\log Z} \rangle = \langle e^{-Ns{\bm e}}\rangle
\end{equation}
where in the last step we used the relation \eqref{min_free_energy_intro}. Now we can use the large deviation form \eqref{mean_Large_dev} and rewrite the above as:
\begin{equation}\label{LD2}
\lim_{n=s/\beta, \beta\to \infty}\langle Z^n  \rangle = \langle e^{-Ns{\bm e}}\rangle = \int d{\bm e} {\cal P}_N({\bm e}) e^{-Ns{\bm e}} \approx \int d{\bm e} R({\bm e}) e^{-N\left(s{\bm e}+{\cal L}({\bm e})\right)}
\end{equation}
which obviously suggests evaluating the integral by the Laplace method as $N\gg 1$, giving
\begin{equation}\label{LD_Legendre}
\lim_{n=s/\beta, \beta\to \infty}\langle Z^n  \rangle  \approx g(s)e^{N\phi(s)}, \quad \phi(s)=-\min_{{\bm e}} (s{\bm e}+{\cal L}({\bm e}))
\end{equation}
We see that the large deviation rate ${\cal L}({\bm e})$ is related by the so-called Legendre transform
to the function $\phi(s)$. Hence, if one  can by independent means find $\phi(s)$, one recovers  ${\cal L}({\bm e})$
by the inverse Legendre transform:
\begin{equation}\label{LD_Legendre1}
{\cal L}({\bm e})=-\left({\bm e}s_*+\phi(s_*)\right), \quad e=-\phi'(s_*)
\end{equation}
 The ability to find the large deviation rate for the minimum of the cost/loss function explicitly hinges on the feasibility to evaluate the above limit  and
to perform the Legendre inversion in a closed-form. Technically, this is a difficult task and  can be rarely successfully performed. It turns out that our problem is one of these rare cases.

In the following section we give brief account of the main results of the paper, first in the Lagrange mutiplier setting, and then in the statistical mechanics approach. Some formulas in the first part have been reported
(without derivation) in our earlier conference proceedings \cite{FT20}.

\section{ A summary and discussion of the main results}
We start with discussing results   obtained via the Lagrange multiplier method.
 Following the standard idea of a constrained minimization, one uses the cost function \eqref{energy} to build the associated Lagrangian ${\cal L}_{\lambda,{\bm s}}({\bm x})=H({\bm x})-\frac{\lambda}{2}({\bm x},{\bm x})$, with real $\lambda$ being the Lagrange multiplier taking care of the spherical constraint. The stationarity conditions $\nabla {\cal L}_{\lambda,{\bm s}}({\bm x})=0$ combined with the spherical constraint yield the equation for the Lagrange multiplier $\lambda$ in the form
	\begin{equation}\label{equlambda}
		{\bm b}^TA(W-\lambda I_N)^{-2}A^T{\bm b}=N
	\end{equation}
 where $W=A^TA$.  Every real root $\lambda$ solving \eqref{equlambda} gives us a Lagrange multiplier corresponding to a stationary point of the cost function, and determines the position $\bm x_{\lambda}$ of the corresponding stationary point.  An important property proved in \cite{Browne1967} is that the order of Lagrange multipliers exactly corresponds to the order of values taken by the cost function at the corresponding  point ${\bm x}$.  Namely, denoting $\cal N$ the total number of Lagrange multipliers, assumed to be distinct and ordered as  $\lambda_1<\lambda_2<\ldots <\lambda_{\cal N}$, such order implies $H({\bm x}_1)<H({\bm x}_j)<\ldots <H\left({\bm x}_{\cal N}\right)$. Thus the minimal loss is always given by ${\cal E}_{min}=H\left({\bm x}_{min}\right)$, where ${\bm x}_{min}$ corresponds to $\lambda_1:=\lambda_{min}$.

 To develop some intuition about the number of solutions of the equation \eqref{equlambda} as a function of the noise variance parameter $\sigma$ it is useful to write another representation of the left-hand side which turns out to be more insightful. To this end, alongside with the Wishart $N\times N$ matrix  $W = A^T A$ we define the associated $M \times M$ matrix $W^{(a)}=AA^T$. Recalling $M\ge N$, the two matrices share the same set of nonzero eigenvalues $\{s_i\}_{i=1}^N$ and $W^{(a)}$ has another $M-N$ eigenvalues equal exactly to $0$. Recalling the spectral decomposition $W^{(a)}=\sum_{i=1}^N s_i \bm v_i\otimes \bm v_i^T$ where $\{\bm v_i\}_{i=1}^M$ are the normalized eigenvectors of $W^{(a)}$ corresponding to its non-zero eigenvalues $s_i, \, i=1, \ldots, N$, and using the associated singular value representation for matrices $A$  allows to rewrite \eqref{equlambda} equivalently as
	\begin{equation}\label{equlambda_new_xi_1}
		\sum\limits_{i=1}^N \frac{s_i(\bm{\xi}^T \bm v_i)^2}{(\lambda-s_i)^2} = \frac{N}{\sigma^2},
	\end{equation}
	where we represented $\bm b$ as $\bm b = \sigma^2 \bm \xi$ via introducing  a mean-zero normally distributed random vector ${\bm\xi}=(\xi_1, \dots , \xi_M)^T$ with the i.i.d. unit variance components:  ${\langle \xi_k \xi_l\rangle=\delta_{kl}}$. Note also that $s_i  (\bm{\xi}^T \bm v_i)^2 \geq 0$ everywhere. It is easy to see that the left-hand side is a positive function of $\lambda$ having a single minimum between every consecutive pair of eigenvalues of $W$, see figure \ref{fig g for N=5} below. This implies there are typically 0 or 2 solutions of eq.~\eqref{equlambda_new_xi_1} (and 1 solution with probability zero at exceptional points) for $\lambda$ between every consecutive pair of eigenvalues, plus two more solutions: the minimal one $\lambda_{min}\in (-\infty, s_1)$ and the maximal one $\lambda_{max}\in ( s_N, \infty)$. Note that the latter two solutions exist for any value of $\sigma\in [0,\infty]$, whereas by changing $\sigma$ one changes the number of solutions available between consecutive eigenvalues. In particular, in the limit of vanishing noise (i.e. $\sigma\to 0$, hence ${\bm b}\to 0$), every stationary point solution for the Lagrange multiplier corresponds to an eigenvalue $s_k$ of the Wishart matrix, with ${\bm x}=\pm {\bm e}_k$ being the associated eigenvectors (hence there are $2N$ stationary points).
	On the other hand when $\sigma \to \infty$ the ratio $N/\sigma^2$ in the right-hand side becomes smaller than the global minimum of the left-hand side in $[s_1,s_N]$. Then  only two stationary points remain: $\lambda_{max}$ and $\lambda_{min}$.

	\begin{figure}
         \includegraphics[width=.75\textwidth]{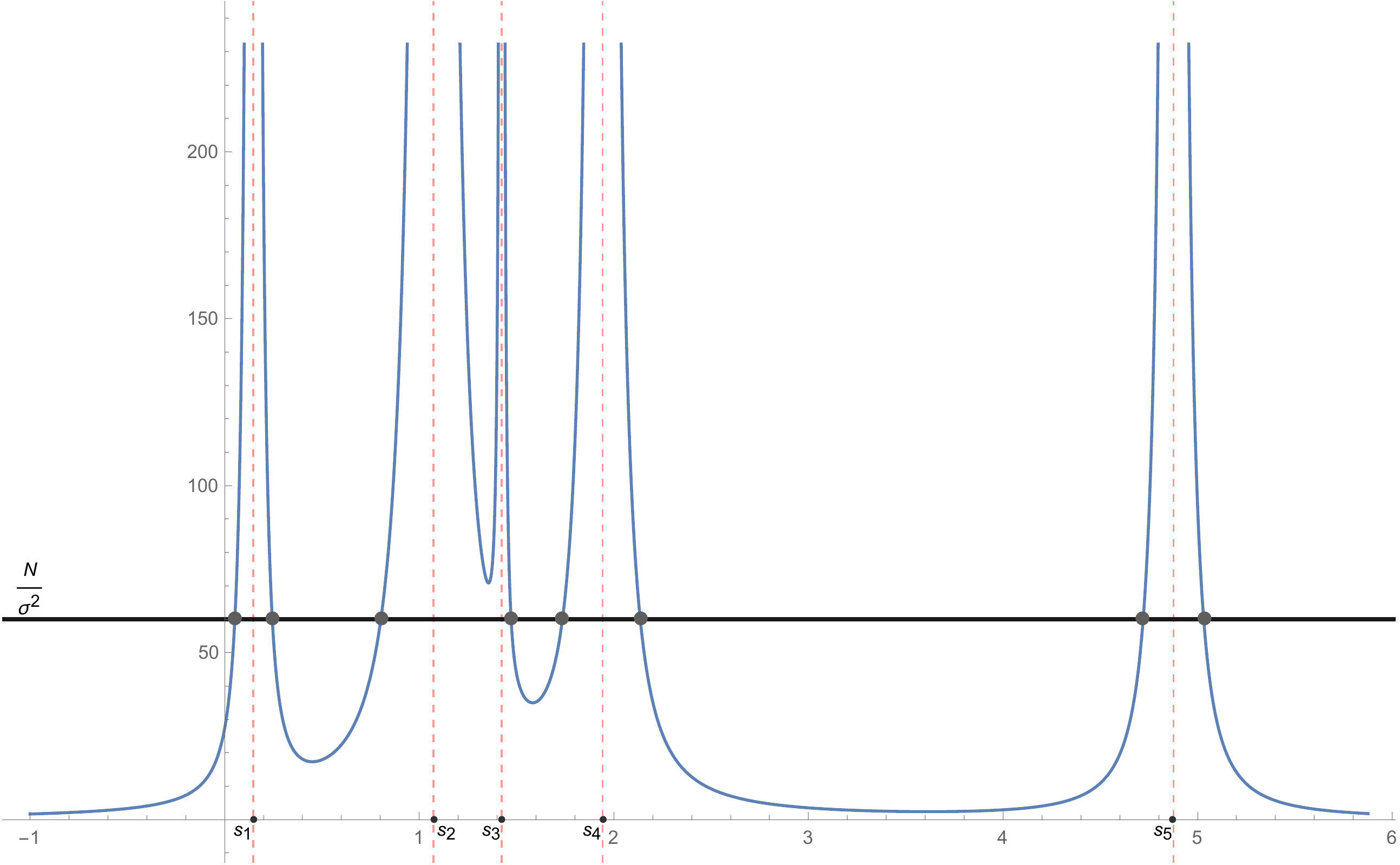}
		\caption{The graph for case $N=5$ representing solution to the equation \eqref{equlambda_new_xi_1}.}
		\label{fig g for N=5}
	\end{figure}

This is an example of the ``gradual topology trivialization" phenomenon, first discussed for a closely related (but different) Gaussian random matrix-based optimization landscape on the sphere treated in \cite{FLD2013}. Obviously, in every particular realization the number of stationary points will change between the two limits, forming a staircase $\mathcal N_\sigma$ as a function of growing $\sigma$. Averaging this staircase over different realizations of both random matrices and the noise one obtains a smooth function $\left\langle \mathcal N\right\rangle_\sigma$, representing the average number of roots vs. $\sigma$ for a fixed $N,M$.  Our goal is to provide the explicit analytical description of such a smooth function using the Kac-Rice approach, and investigate its asymptotics as $N\to \infty$. Note that the methods of \cite{FLD2013} can not be immediately applied to the present case since consideration there	essentially used that the cost functions was Gaussian-distributed mean zero random field, whereas our cost function \eqref{energy} is always non-negative and given by the sum of squared Gaussian variables.

 In what follows we find it convenient to redefine the control noise parameter $\sigma$ and use $\delta$ defined by $\delta=\frac{1}{2}\log{(1+\sigma^2)}$ (equivalently, $\sigma^2=e^{2\delta}-1$).
Then using the Kac-Rice method we show in the Section  (\ref{The average number of stationary points}) that the mean number of real Lagrange multipliers (hence stationary points) is given by the sum of two contributions:
	\begin{equation}\label{tot}
		\left\langle {\cal N} \right\rangle_\delta=\left\langle {\cal N}\right\rangle_\delta^{^+}+
		\left\langle {\cal N}\right\rangle_\delta^{^-}
	\end{equation}
	The first contribution counts the mean number of positive Lagrange multipliers $\lambda\in [0,\infty)$ and  can be represented as
\begin{equation}\label{numberpos}
\left\langle {\cal N}\right\rangle^{^+}_\delta= \int_0^{\infty} p(\lambda)\, d\lambda
\end{equation}
where the associated density $p(\lambda)$ given by
	\begin{equation}\label{main}
		 p(\lambda)= \sqrt{\frac{N}{\pi}}\frac{e^{-\frac{M+N-1}{2}\delta}}{\sqrt{\sinh{\delta}}}\left\langle\rho_N(\lambda)\right\rangle \sqrt{\lambda}
		\int_{-\infty}^{\infty}e^{\frac{M-N}{2}t} \, e^{-N\frac{\lambda}{2}\frac{\cosh{t}-\cosh{\delta}}{\sinh{\delta}}}\,dt
		\end{equation}
	where $\left\langle\rho_N(\lambda)\right\rangle=\left\langle\sum_{k=1}^N\delta(\lambda-s_k)\right\rangle$ is the mean eigenvalue density of the Wishart ensemble \eqref{Wishart}, normalized as $\int  \left\langle\rho_N(\lambda)\right\rangle \,d\lambda=N$.
The second contribution counts the mean number of negative Lagrange multipliers and is given by
	\begin{align}
		\langle \mathcal N \rangle^{^-}_\sigma =& 2\sqrt{\pi} \frac{ (N-1)!(M-1)!}{\Gamma\left(\frac N 2\right)\Gamma \left( \frac{M}2 \right)} \frac {\sigma^{M-N-1}}{\left(1+ \sqrt{1+\sigma^2}\right)^{M-N} \left(2\sqrt{1+\sigma^2}\right)^{N-1}}  \notag\\
		&\times  \sum\limits_{k=0}^{N-1}  \left(\frac{\sigma^2}{1+ \sqrt{1+\sigma^2}}\right)^{k+1} \frac{1}{\Gamma(N-k)\Gamma\left(\frac{M-N+1}{2}+k+1\right)} \notag\\
		&\times F\! \left( M-N+k+1, \frac{M-N+1}{2},\frac{M-N+1}{2}+k+1, \frac{1-\sqrt{1+\sigma^2}}{1+ \sqrt{1+\sigma^2}} \right) \label{subleading}
	\end{align}
	where $F(a,b;c;z) \equiv _2 \!\! F_1 (a,b;c;z)$ is the ordinary hypergeometric function.
As in our problem only the smallest Lagrange multiplier may take negative values, the value $\left\langle {\cal N}_\delta \right\rangle^{^-}$ always satisfies $0<\left\langle {\cal N}_\delta \right\rangle^{^-}\leq 1$.

For finite values of $N,M$ the resulting expressions for $\langle \mathcal N \rangle^{^\pm}_\sigma$   can be easily plotted. For example, we show below $\langle \mathcal N \rangle^{^-}_\sigma$ for different $N$ and a fixed ratio $\alpha=M/N=1.5$: 	
	\begin{figure}[h!]
		\center
\includegraphics[width=.75\textwidth]{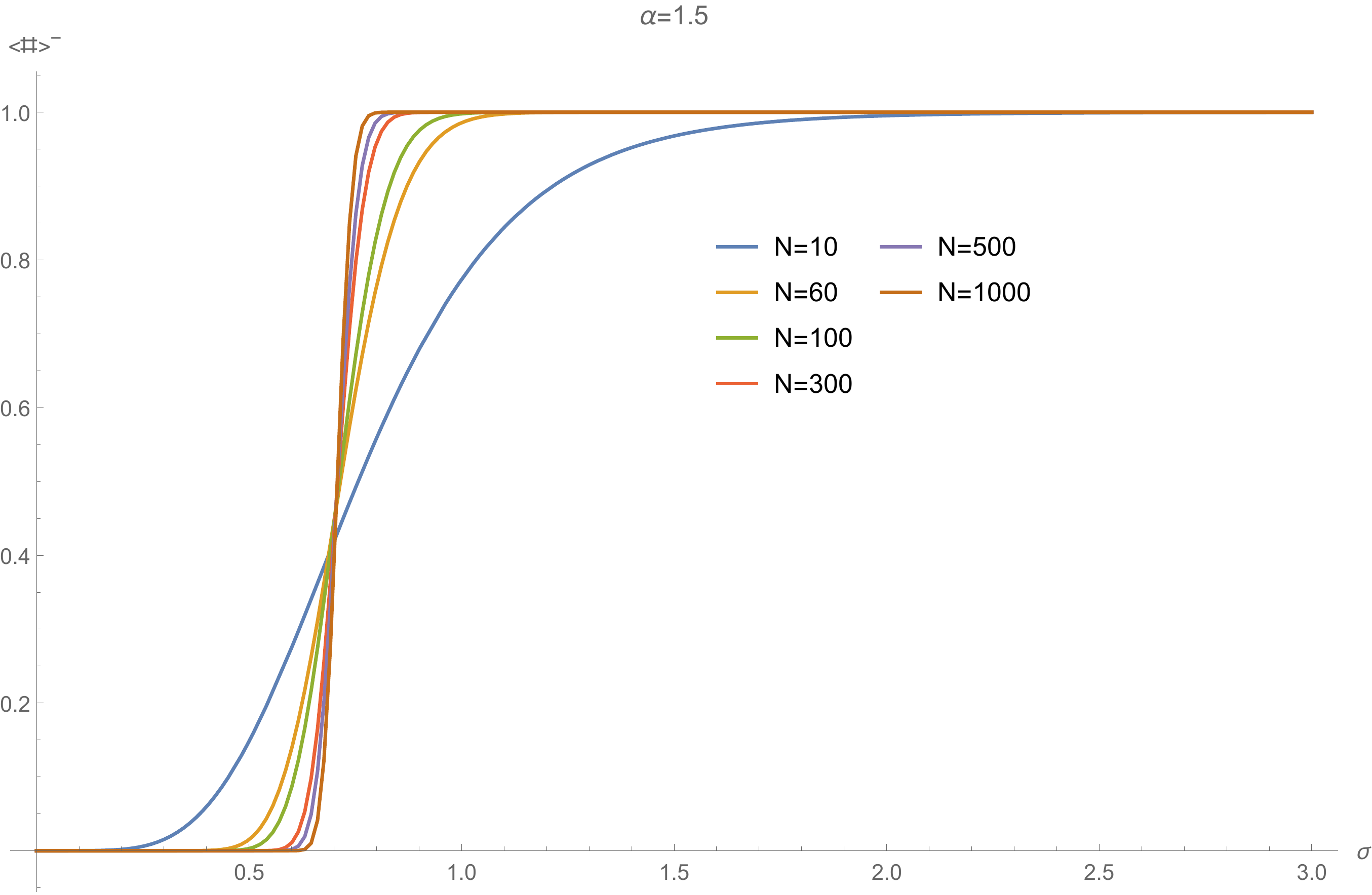}
		\caption{Values of $\langle \mathcal N \rangle^{^-}_\sigma$ from \eqref{subleading} plotted vs. $\sigma$ for matrices with various sizes $N$.}
		\label{av Num minus pic}
	\end{figure}
	The above formulas are valid for any finite $N<M$ and can be further used for asymptotic evaluation in the limit
 $N\to \infty$ such that $M/N=\alpha>1$ is kept fixed. The limiting form of $\left\langle \mathcal N_\sigma^{^-} \right\rangle$ turns out to be simple:
\begin{equation} \label{47}
		\lim_{N\rightarrow\infty}  \left\langle \mathcal N_\sigma^{^-} \right\rangle  = \begin{cases}
			0, &\mbox{ if } \sigma < \sqrt{\alpha - 1}\\
			1, &\mbox{ if } \sigma > \sqrt{\alpha -1}
		\end{cases}
	\end{equation}

 and such behaviour is indeed evident from Fig.\ref{av Num minus pic}. The asymptotic behaviour of
 $\left\langle \mathcal N_{\sigma}^{+}\right\rangle$ is much more interesting and reflects the topolgy trivialization.

 Qualitatively it is not difficult to see from (\ref{equlambda_new_xi_1}) that as $N\gg 1$ the trivialization gradually must happen on the scale $\sigma^2\sim 1/N$ as only for such values the left-hand side is of the same order as the right-hand side for a generic $\lambda \in [s_k,s_{k+1}]$ (in our normalization the typical distance is $|s_k-s_{k+1}|=O(N^{-1})$). To describe this trivialization process quantitatively via the mean counting function we recall that in the above limit the mean density of Wishart eigenvalues is given by the Marchenko-Pastur \cite{MP} expression:
	\begin{equation}\label{MP}
		\frac{1}{N}\left\langle\rho_N(\lambda)\right\rangle \to \rho_{MP}(\lambda)=\frac{2}{\pi}\frac{1}{\left(\sqrt{s_{+}}-
			\sqrt{s_{-}}\right)^2}\,\frac{\sqrt{(\lambda-s_{-})(s_{+}-\lambda)}}{\lambda}, \quad s_{-}\leq \lambda\leq s_{+},
	\end{equation}
	and zero for $\lambda \notin (s_{-} s_{+})$,
where $s_{\pm}=(\sqrt{\alpha}\pm 1)^2$ are the positions of the so-called spectral edges.
	Using this it is possible to show that if $0<\delta<\infty $ is fixed
	and $N\to \infty$ then the number of solutions is minimal possible: $\lim_{N\to \infty}\left\langle {\cal N} \right\rangle_\delta=2$. Following  \cite{FLD2013} it is natural to call this phenomenon the complete ''landscape topology trivialization'', as only a single minimum and a single maximum remain in the cost function profile. For a non-trivial cost landscape to survive in the limit $N\to \infty$ one needs to consider the noise scaled as  $\delta\sim N^{-1}$. Introducing a scaling parameter $\gamma=\delta N/4$ and considering it to be finite when $N\to \infty$ then yields:
	\begin{equation}\label{mainlim}
		\lim_{N\to \infty} \frac{\left\langle {\cal N}\right\rangle^{^+}_\gamma}{N}=\frac{2}{\pi}\int_{s_{-}}^{s_+}\, \frac{\sqrt{(\lambda-s_{-})(s_{+}-\lambda)}}{\lambda}\, e^{-\frac{\gamma}{\lambda}(\lambda-s_{-})(s_{+}-\lambda)}\,d\lambda
	\end{equation}
	We see that in this regime the number of stationary points is always of the order of $N$, and decreases with increasing parameter $\gamma$. In particular, one can further find that asymptotically for large $\gamma\gg 1$
	\begin{equation}\label{mainlimasy}
		\lim_{N\to \infty} \frac{\left\langle {\cal N}\right\rangle_\gamma^{^+}}{N} \Biggr|_{\gamma\gg 1}\approx \frac{1}{2\sqrt{\pi}\gamma^{3/2}}
	\end{equation}
	The above formula suggests that for $\gamma \sim N^{2/3}$ (which in the original parametrization corresponds to $\delta\sim N^{-1/3}$)
	one should expect the number of stationary points to be of the order of unity. Moreover, for any fixed $\alpha>1$ we should expect the lowest Lagrange multiplier to be still typically positive in this regime as $\lim_{N\rightarrow \infty } \left\langle \mathcal N \right\rangle^{^-}_{\delta\sim N^{-1/3}} = 0$, as follows from \eqref{47}.
	To get the correct counting formula in this regime one also can not use
	simply the Marchenko-Pastur density \eqref{MP}, but needs to use the ``edge scaling'' for the eigenvalue density. Indeed, the corresponding integral turns out to be dominated by the contribution from the
Marchenko-Pastur spectral edges $s_\pm$ and
	\begin{equation}\label{rho edge}
		\left\langle \rho (\lambda)\right\rangle \rightarrow  \left(\frac{s_+-s_-}{4 N s_\pm^2}\right)^{\frac 13} \rho_{edge} (\xi),
	\end{equation}
	with the edge  density $\rho_{edge} (\xi)$ given in terms of Airy function $\Ai(\xi) = \frac 1{2\pi i} \int\limits_\Gamma dv\, e^{\frac{v^3}{3}-v\xi}$ as (see e.g. \cite{ForrLD})
	\begin{align}\label{edgedens}
		\rho_{edge} (\xi) =& [\Ai' (\xi)]^2 - \xi [\Ai (\xi)]^2 + \frac 12 \Ai (\xi) \left( 1 - \int\limits_\xi^\infty \Ai (y) dy \right)
	\end{align}
In this scaling we find that the mean number of Lagrange multipliers/stationary points is given by
	\begin{equation}\label{edgecontrib}
		\lim_{N\rightarrow \infty } \left\langle \mathcal N \right\rangle^{^+}_\omega = 2 \int\limits_{-\infty}^{\infty} \left[ e^{-\frac{\omega^3}{3s_-} + \frac{\omega\xi}{s_-^{1/3}}} +e^{-\frac{\omega^3}{3s_+} + \frac{\omega\xi}{s_+^{1/3}}} \right] \rho_{edge} (\xi) d \xi,
	\end{equation}
	where we introduced a new parameter
	\begin{equation*}
		\omega= N^{\frac 13} \delta \left( \dfrac{s_+-s_-}{4}  \right)^{\frac 23} \in [0,\infty].
	\end{equation*}
	Taking the limit $\omega \rightarrow \infty$ in \eqref{edgecontrib} gives $2$ as the limiting value, which confirms the tendency for any fixed noise variance $\sigma^2>0$ to have typically only two stationary points in the cost function landscape on the sphere (which is the minimal possible number), corresponding to a minimum and a maximum.

Next we show that the density $p(\lambda)$ of positive Lagrange multipliers
in the range $0<\lambda<s_{-}$ has  the following large deviations form
	\begin{equation}\label{LDsmallest}
		p(\lambda_{min}) \sim e^{-\frac N2 \Phi (\lambda_{min})}, \quad 0<\lambda<s_{-}
	\end{equation}
where $\Phi (\lambda) = L_1 (\lambda) + L_2 (\lambda) + const$, with
\be
const= \frac{\alpha +1}{2} \log(1+\sigma^2) + 2(\alpha - 2) \log{\frac{1}{2\sqrt{\alpha}}},
\ee		
whereas
\begin{equation}\label{L1}
L_1 (\lambda) = (\alpha - 1)\left[ \frac{\sqrt{ \lambda^2+\kappa^2}}{\kappa} - \log \left( \kappa + \sqrt{ \lambda^2 + \kappa^2} \right) -  \lambda \frac{\sqrt{(\alpha - 1)^2 + \kappa^2}}{(\alpha -1 ) \kappa}   \right],
\end{equation}
where we defined $\kappa = (\alpha-1)\sinh \delta$ and
\begin{align}
		L_2 (\lambda) =& - \sqrt{(\lambda - s_-)(\lambda - s_+)} - 2\log (\alpha + 1 - \lambda + \sqrt{(\lambda - s_-)(\lambda - s_+)}) \notag \\
		&+ 2 (\alpha -1) \log(\lambda + \alpha -1 + \sqrt{(\lambda - s_-)(\lambda - s_+)}). \label{L_2}
	\end{align}
Since for $N\to \infty$ typically only the smallest Lagrange multiplier $\lambda_{min}:=\lambda_1$ can belong
to the range $0<\lambda<s_{-}$,
we interpret \eqref{LDsmallest} as providing the large deviation form for the probability density of
$\lambda_{min}$. This implies that by finding such  $\lambda_*$ which minimizes $\Phi (\lambda)$ gives the most probable value of the minimal Lagrange multiplier. We are able to show that
	\begin{equation}
		\lambda_\ast = \left(\sqrt \alpha - \sqrt{1+\sigma^2}\right) \left( \sqrt{\alpha}-\frac 1{\sqrt{1+\sigma^2}}  \right), \quad \alpha>1
	\end{equation}
Note that $\lambda_\ast$ changes its sign at $\alpha=1+\sigma^2>1$, i.e. precisely the same value of
$\alpha$ when the mean number $\left\langle \mathcal N_\sigma^{^-} \right\rangle$ of negative Largange multipliers
in \eqref{47} changes from $0$ to $1$, as should be expected on consistency reasons. Finally, we find that the mean cost/loss function at $\lambda_*$ is given by
	\begin{equation}\label{minmeanvalue}
	{\bm e}_{min}=\lim\limits_{N\rightarrow \infty} \frac{\langle \mathcal E_{\lambda_\ast} \rangle}{N} = \frac 12 \left( \sqrt{\alpha (1+\sigma^2)} -1 \right)^2,
	\end{equation}
giving the asymptotic minimal value for the cost. Note, that in the RMT-based derivation above we assumed $\alpha>1$. However, one can use a different method (see below) to show that the formula \eqref{minmeanvalue} is actually valid beyond that constraint and that ${\bm e}_{min}$ remains positive as long as $1/(1+\sigma^2):=\alpha_c<\alpha<1$. For any $\alpha<\alpha_c$ one finds in contrast ${\bm e}_{min}=0$.
 The only possible interpretation is that not only overcomplete systems of linear equations (with $\alpha>1$) are typically incompatible on the sphere, but incompatibility extends also to a large range of undercomplete systems with $\alpha<1$, and only when
$\alpha=M/N$ approaches $\alpha_c$ from above the system starts to be compatible, ensuring the cost to take its minimal possible value zero. From this viewpoint it is natural to refer to the value $\alpha_c=1/(1+\sigma^2)$ as the {\it (in)compatibility threshold}.

Note that in deriving \eqref{minmeanvalue} in the RMT-based way we relied on an assumption about the non-fluctuating (self-averaging) nature of the resolvents (like one in the left-hand side of \eqref{equlambda}) outside spectrum for $\lambda < s_-$. The fact that the value \eqref{minmeanvalue} coincides with the typical minimum of the cost function which can be independently found by the Statistical Mechanics approach justifies validity of our approach.   	

 In the section \ref{Large deviations for the minimal cost} we will demonstrate how the Statistical Mechanics procedure works in the present case, yielding the following Large Deviation Rate function for the minimal cost:
\begin{equation}\label{LD mincost_main}
{\cal L}({\bm e})={\bm e}v-\frac{1}{2v}-\frac{\alpha-1}{2}\log{(2{\bm e})}+\frac{\alpha}{2}\log{\left(\alpha(1+\sigma^2)\right)}+\log{v}-\alpha\log{(1+v)}
\end{equation}
where $v$ for a given ${\bm e}$ should be found by solving the following cubic equation:
\begin{equation}\label{cubic}
\sigma^2v^3+v^2(\sigma^2-1)+v\left(\frac{(\alpha-1)(1+\sigma^2)}{2{\bm e}}-1\right)-\frac{1+\sigma^2}{2{\bm e}}=0
\end{equation}
To choose the correct root it is helpful to notice that for the special value ${\bm e}_*:=\frac{1}{2}\left(\sqrt{\alpha(1+\sigma^2)}-1\right)^2$ the relevant solution of the cubic equation \eqref{cubic} is given by $v_*:=1/\sqrt{{2\bf e}_*}$. This indeed can be checked directly. Note that this special value
${\bm e}_*$ is nothing else but the typical minimal cost ${\bm e}_{min}$ from \eqref{minmeanvalue}.
It minimizes the function  ${\cal L}({\bm e})$ in \eqref{LD mincost_main}, and moreover ${\cal L}({\bm e}_*)=0$ as expected and can easily checked from  \eqref{LD mincost_main} by substituting there the values ${\bm e}_*$ and $v_*$.

Recall that at $\sigma=0$ the minimal cost in our problem must be given simply by the minimal eigenvalue of the Wishart matrix.  The corresponding Large Deviation Rate in such a case has been obtained a few years ago in \cite{KazCasLargeDev} using Coulomb Gas techniques, and fully rigorously in  \cite{ForrLD}. It is interesting to see how this result is reproduced in the present approach. As $\sigma^2=0$ the equation \eqref{cubic} is reduced to the quadratic
\begin{equation}\label{quadratic}
v^2+v\left(1-\frac{\alpha-1}{2{\bm e}}\right)+\frac{1}{2{\bm e}}=0
\end{equation}
with the relevant solution for $\alpha>1$
\begin{equation}\label{quadratic}
v=-\frac{1}{2}+\frac{\alpha-1}{2\epsilon}-\frac{1}{2\epsilon}\sqrt{\epsilon^2-2\epsilon(\alpha+1)+(\alpha-1)^2}, \quad \epsilon=2{\bm e}
\end{equation}
In particular, for $ \epsilon=(\sqrt{\alpha}-1)^2:=\epsilon_*$ which corresponds to the typical minimal eigenvalue of the Wishart matrix, we have $v=1/\sqrt{\epsilon_*}:=v_*$. Introducing the deviation from the typical minimal eigenvalue $x= \epsilon_*-{\epsilon}$ and using the notation $\Delta=4\sqrt{\alpha}$ one may notice that $\sqrt{\epsilon^2-2\epsilon(\alpha+1)+(\alpha-1)^2}=\sqrt{x(x+\Delta)}$. Substituting $v$ from \eqref{quadratic}
into \eqref{LD mincost_main} one after some algebra can bring it to the form:
\begin{equation}\label{LD mincost_eigenv}
{\cal L}(x)=-\frac{1}{2}\sqrt{x(x+\Delta)}-\frac{\alpha-1}{2}\log{\frac{\epsilon_*-x}{\epsilon_*}}+
(\alpha-1)\log{\left(1+\frac{2\sqrt{x(x+\Delta)}}{\Delta\sqrt{\epsilon_*}}\right)}+
2\log{\left(\frac{\sqrt{x+\Delta}-\sqrt{x}}{\sqrt{\Delta}}\right)}
\end{equation}
which is exactly the form given in \cite{KazCasLargeDev}.
\begin{figure}[h!]
		\begin{minipage}{.5\textwidth}
			\centering
\includegraphics[width=.9\textwidth]{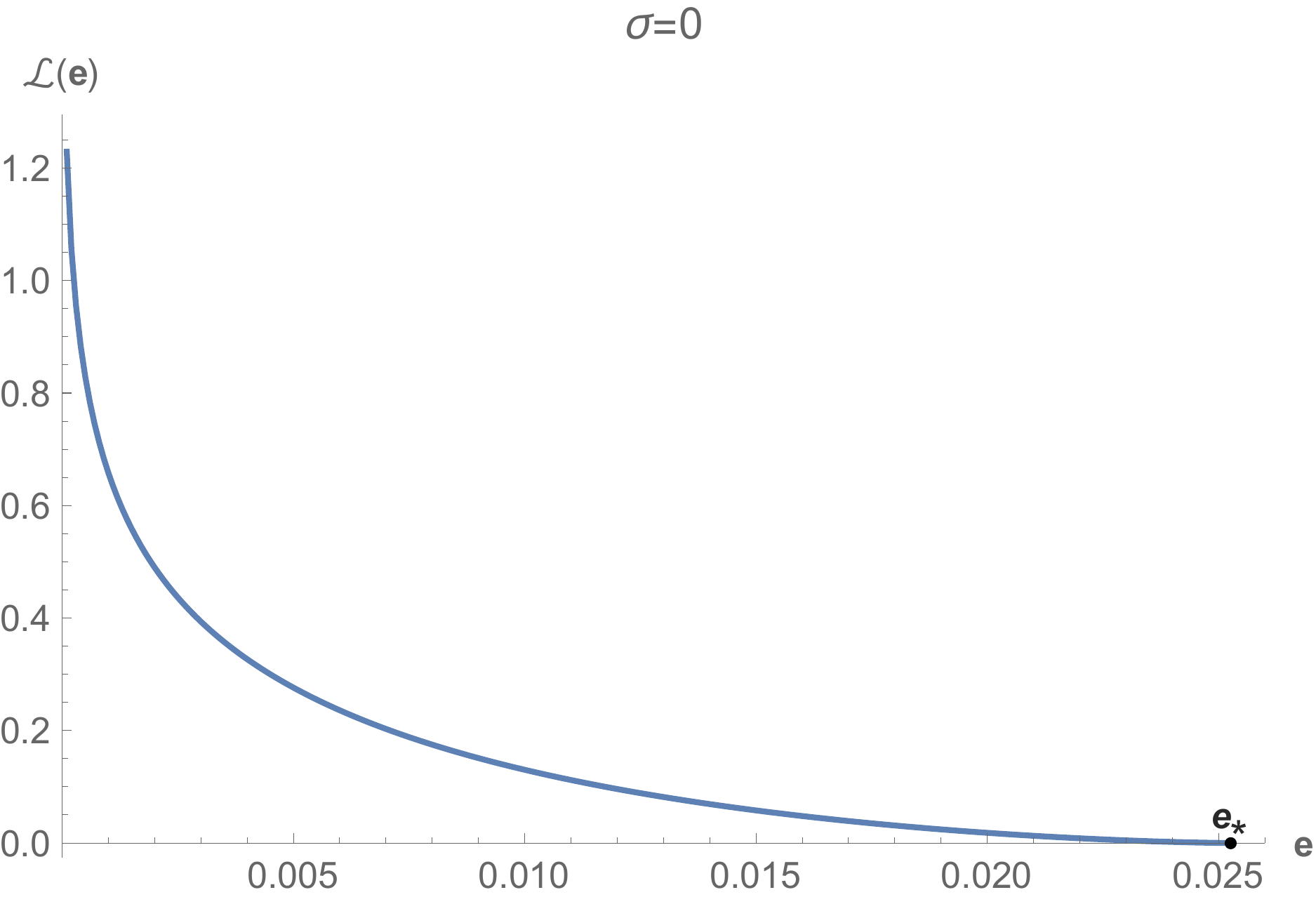}
			\caption{The Large Deviation Rate ${\cal L}({\bm e})$ for~${\alpha=1.5}$ and $\sigma=0$}
			\label{LDR sigma=0}
		\end{minipage}
		\begin{minipage}{.5\textwidth}
			\centering
\includegraphics[width=.9\textwidth]{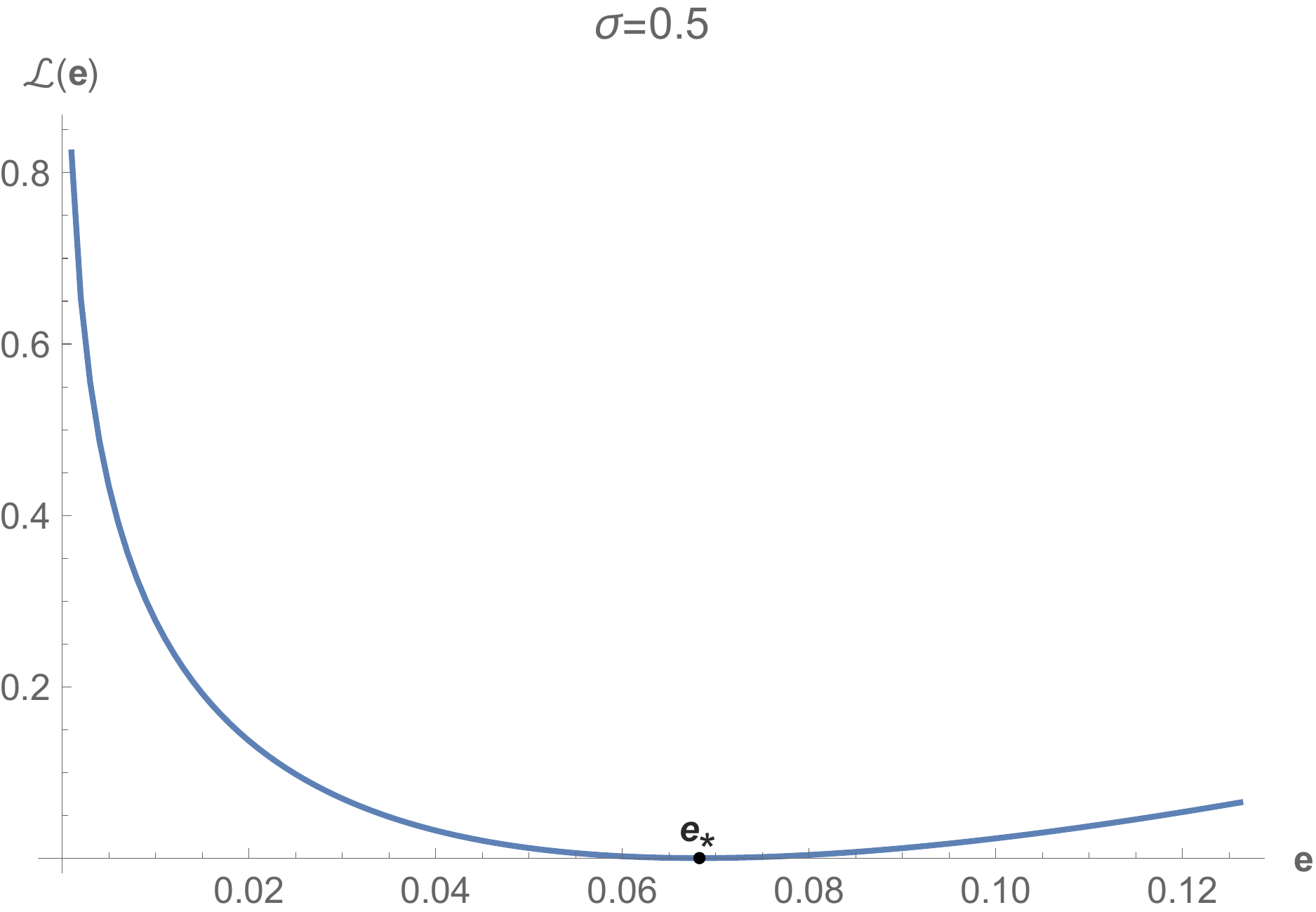}
			\caption{The Large Deviation Rate ${\cal L}({\bm e})$ for $\alpha=1.5$ and $\sigma=0.5$}
			\label{LDR sigma=0.5}
		\end{minipage}
	\end{figure}

Note that for $\sigma=0$ this large deviation rate makes sense only for $x>0$, i.e. ${\bm e}<{\bm e}_*$. In fact it is well-known that for ${\bm e}>{\bm e}_*$ the large deviation probability for minimal eigenvalue  behaves in a different way:  $ {\cal P}_N({\mathcal E}_{min})\sim e^{-N^2{\cal L_+}({\bm e})}$,
 reflecting Coulomb repulsion between eigenvalues of random matrices, see  e.g.~\cite{DeanMaj2006}.

 As long as $\sigma>0$ the formula \eqref{LD mincost_main} is expected to be valid for some range of $\epsilon$ around the mean/typical value, that is also for some values ${\bm e}>{\bm e}_*$,
 and indeed it is easy to plot it using \eqref{LD mincost_main}, see the figures~\ref{LDR sigma=0} and \ref{LDR sigma=0.5}. Its actual range of validity is however not clear to us at the moment, and clarifying it remains an unsolved issue. To this end see the work \cite{DemZeit} which clarified similar issues for a simpler but related replica calculation in \cite{FLD2013}.

 \subsection{Conclusions and Perspectives}
 Our present work provides a reasonably detailed picture of the simplest least-square random cost landscape. From mathematical viewpoint it combines both rigorous and conjectural results, the latter being those obtained in the framework of the replica trick. As such, they call for a proper justification by mathematically rigorous methods.
 In particular, as long as $\sigma>0$ the formula \eqref{LD mincost_main} is expected to be valid for some range of $\epsilon$ around the mean/typical value, that is also for some values ${\bm e}>{\bm e}_*$,
 and indeed it is easy to plot it using \eqref{LD mincost_main}, see the figures~\ref{LDR sigma=0} and \ref{LDR sigma=0.5}. Its actual range of validity is however not clear to us at the moment, and clarifying it remains an unsolved issue. To this end  it is also worth mentioning that many questions for a simpler optimization problem on the sphere studied originally in  \cite{FLD2013} by a combination of rigorous Kac-Rice and heuristic replica approaches were subsequently successfully put on the firm mathematical ground in the series of papers \cite{DemZeit,Kivimae,Sosoe20,Baiketal21,Belius21}. We hope some of those techniques can be also useful in the context of present model as well.

 From a more general perspective, our work suggests a possibility of studying more general random least-square optimization landscapes. In the forthcoming work we plan to address  such landscapes arising from considering systems of $M$ random quadratic equations in $N$ variables on a $N-$sphere, aiming to find the associated compatibility threshold. Such problem seems to be not much studied, though some related questions were addressed recently in special cases motivated by the so-called "phase retrieval" problem, see \cite{phaseretrieval1}.
 Our partial results in this direction can be found in the Phd thesis of the second author
 \cite{PhD_Rashel}.  In particular, applying to that problem the replica-based Statistical Mechanics analysis
 based on the Appendix (\ref{partion_fun_evaluation}) of the present paper we are able to show
 that finding the minimal cost requires, in a broad range of parameters, to use the Full Replica Symmetry Breaking scheme due to Parisi \cite{Parisi79}. In contrast, the same analysis shows that the problem considered
 in the present paper can indeed be solved in the Replica Symmetric Ansatz, further justifying with the hindsight
 the approach adopted here.

\ack
 The research of the first author was supported by the EPSRC Grant EP/V002473/1 {\it Random Hessians and Jacobians: theory and applications}.
\\[2ex]

\section{Derivation of the main results} \label{Derivation_Chap1}
	\subsection{The average number of stationary points}\label{The average number of stationary points}
	 Taking the gradient of the  Lagrangian ${\cal L}_{\lambda,{\bm s}}({\bm x})=H({\bm x})-\frac{\lambda}{2}({\bm x},{\bm x})$ and setting it to zero implies the following stationary condition
	\begin{equation}
		{A^T\left[A{\bm x}-{\bm b}\right]-\lambda {\bm x}=0},
	\end{equation}
which together with the normalization condition ${\bm x}^2=N$ gives the system of $N+1$ equations for unknowns $x_1,\ldots,x_N$ and $\lambda$.	 Then,  the Kac-Rice method applied to counting Lagrange multipliers following the approach of \cite{Fyo16} implies that the mean total number $\left\langle {\cal N}\right\rangle$ of stationary points of the Lagrangian, when $\bm x$ is restricted to an $N$-sphere can be written as
	\begin{equation}\label{KacRice}
		\langle {\cal N} \rangle = \left\langle \int d\lambda \int \delta \left[ A^T A {\bm x} - A^T {\bm b}-\lambda {\bm x} \right] \delta \left({\bm x}^2 - N  \right)  \left| \det
		\begin{pmatrix}
			A^TA-\lambda I_N & {\bm x}\\
			-2{\bm x}^T & 0
		\end{pmatrix}
		\right| d{\bm x} \right\rangle_{A,{\bm b}}
	\end{equation}
	 We begin with averaging the above over the Gaussian random vector $\bm b$  which is straightforward due to the presence of the corresponding $\delta-$function:
	\begin{equation}\label{23}
	\left\langle \delta (\bm u - A^T \bm b)\right\rangle  = \frac{\exp{{-\frac{1}{2\sigma^2}\bm u^TW^{-1}\bm u}}}{(2\pi \sigma^2)^\frac N2\det^{\frac 12}W}, \quad \bm u = \left( W-\lambda I_N \right){\bm x},
	\end{equation}
where $W=A^TA$ is the familiar Wishart matrix. Our next step is to perform integration over $\bm x$ in  \eqref{KacRice}. We achieve this by remembering the orthogonal invariance of the Wishart density \eqref{Wishart}, and exploiting the ensuing rotational invariance of the integrands with respect to orthogonal transformations ${\bm x}\to O{\bm x}$. Again following  \cite{Fyo16} we can replace ${\bm x}=\sqrt{N}{\bm e}_1$, with ${\bm e}_1:=(1,0,\ldots,0)$ in the exponential and determinant in the integrand. Then with the help of the identity
\begin{align}\label{ident_A}
			\int_{\mathbb R^N} d{\bm x} \, \delta ({\bm x}^2 - N) f({\bm x}^2)=\frac{2\pi^{N/2}}{\Gamma\left(\frac N 2\right)}  N^{\frac{N-2}2} f(N)
		\end{align}
we perform the integration over ${\bm x}$ explicitly arriving at
	\begin{align}
		&\langle {\cal N}  \rangle = \frac{\mathcal K_N N}{(2\pi\sigma^2)^{\frac N2}}\int d\lambda  \int  \frac{dW \mathcal P_N(W)}{\det^{\frac 12}W } e^{-\frac{N}{2\sigma^2}\left[(W-\lambda I_N)^2W^{-1}\right]_{11}} \left| \det (\widetilde{W}-\lambda I_{N-1})
		\right| \label{without x}\\
		&\mbox{with } {\cal K}_N=\dfrac{2\pi^{N/2}}{\Gamma\left(\frac N 2\right)}  N^{\frac{N-2}2} \mbox{ and } W=\begin{pmatrix} w & {\bm v}^T\\ {\bm v} & \widetilde{W}\end{pmatrix}\mbox{, where $\widetilde{W}$ is Wishart $N-1 \times N-1$.} \notag
	\end{align}
Further writing  $[W^{-1}]_{11} = \frac 1{w-\bm v^T \widetilde{W}^{-1} \bm v}$ we have
	\begin{equation}\label{26}
		\left[(W-\lambda I_N)^2W^{-1}\right]_{11} =w-2\lambda+\frac{\lambda^2}{w-\bm v^T \widetilde{W}^{-1} \bm v}.
	\end{equation}
and note that with respect to our reparametrization of positive definite $W$ the Schur complement formulas imply that $w$, $\widetilde{W}$ and  $w-\bm v^T \widetilde{W}^{-1}\bm v$ are positive definite, further implying the change of the measure and of the determinant as
	\begin{align}
		dW &=   \theta(w) \theta(w-\bm v^T \widetilde{W}^{-1}\bm v)\,dw\, d\bm v\, d\widetilde{W}\\
		\det W &= \det \widetilde{W} \det (w - \bm v^T \widetilde{W}^{-1}\bm v) = (w - \bm v^T \widetilde{W}^{-1}\bm v)  \det \widetilde{W}
	\end{align}
where we used the Heaviside step function $\theta(x)=1$ for $x>0$ and zero otherwise.  These properties, together with the explicit representation of the joint probability density for the Wishart ensemble \eqref{Wishart} allows us to present \eqref{without x} in the form
	\begin{align}
		\langle {\cal N}  \rangle =& \frac{\mathcal K_N N C_{N,M}}{(2\pi\sigma^2)^{\frac N2}}\int d\lambda  \iiint  e^{-\frac N2 \Tr \widetilde{W}}e^{-\frac N2 w }  [w - \bm v^T \widetilde{W}^{-1}\bm v]^{\frac{M-N-2}2} \left[ \det \widetilde{W}  \right]^{\frac{M-N-2}2} \notag\\
		&\times  \theta(w) \theta(w-\bm v^T \widetilde{W}^{-1}\bm v)\,dw\, d\bm v\, d\widetilde{W} \notag \\
		&\times \exp {{-\frac{N}{2\sigma^2}\left( w-2\lambda+\frac{\lambda^2}{w-\bm v^T \widetilde{W}^{-1} \bm v}\right)}} \left| \det (\widetilde{W}-\lambda I_{N-1})
		\right| \label{29}
	\end{align}
	Now lets consider only integration over $w$ and $\bm v$:
	\begin{equation}
		\mathcal I = \int\limits_{\mathbb R^{N-1}} \!\!d\bm v \int\limits_0^\infty dw \,\theta(w-\bm v^T \widetilde{W}^{-1}\bm v)  [w - \bm v^T \widetilde{W}^{-1}\bm v]^{\frac{M-N-2}2}  e^{-\frac N2 w (1+\frac 1{\sigma^2}) -\frac {N\lambda^2}{2\sigma^2} \left(w-\bm v^T \widetilde{W}^{-1} \bm v \right)^{-1}}
	\end{equation}
	Introducing a change of variables $w \rightarrow q = w-\bm v^T \widetilde{W}^{-1} \bm v$ yields
\begin{align*}
		\mathcal I 	=& \int\limits_{\mathbb R^{N-1}} d\bm v \, e^{- \frac N2 \left(1+\frac 1{\sigma^2}\right) \bm v^T \widetilde{W}^{-1}\bm v} \int\limits_0^\infty dq \, q^{\frac{M-N-2}2}  e^{-\frac N2  \left(1+\frac 1{\sigma^2}\right)q -\frac {N\lambda^2}{2\sigma^2} \frac 1q}\\\intertext{and after integrating out $\bm v$  and re-scaling $q \rightarrow \frac{|\lambda|}{\sqrt{1+\sigma^2}} q$ one gets}
	\mathcal I 	=& \frac{(2\pi)^{\frac{N-1}2}}{\det^{\frac 12}\!\! \left[N\left(1+\frac 1{\sigma^2}\right) \widetilde{W}^{-1}\right]} \cdot \left( \frac{|\lambda|}{\sqrt{1+\sigma^2}}  \right)^{\frac{M-N}{2}} \int\limits_0^\infty dq \, q^{\frac{M-N-2}2}  e^{-\frac {N|\lambda|}{2\sigma^2}\sqrt{1+\sigma^2}  \left(q+\frac 1q\right)}\\
		=&\left( \frac{2\pi\sigma^2}N \right)^{\frac{N-1}2}  \frac{\det^{\frac 12} \widetilde{W} }{(1+\sigma^2)^{\frac{M+N-2}{4}}} |\lambda|^{\frac{M-N}{2}}  \int\limits_0^\infty dq \, q^{\frac{M-N-2}2}  e^{-\frac {N|\lambda|}{2\sigma^2}\sqrt{1+\sigma^2}  \left(q+\frac 1q\right)}
	\end{align*}
	which allows to express $\langle \mathcal N  \rangle$ as
	\begin{align}
		\langle \mathcal N  \rangle
		=& \frac{N \mathcal K_N C_{N,M} }{N^{\frac{N-1}{2}}} \frac{1}{\sqrt{2 \pi \sigma^2}} \frac 1{(1+\sigma^2)^{\frac{M+N-2}{4}}} \notag\\
		&\times \int\limits_{-\infty}^\infty d\lambda\, e^{\frac{N\lambda}{\sigma^2}} |\lambda|^{\frac{M-N}2} \int\limits_0^\infty dq \,q^{\frac{M-N-2}{2}} e^{-\frac{N}{2\sigma^2}|\lambda|\sqrt{1+\sigma^2}(q+\frac 1q)} \times \Phi_{N,M} (\lambda),\label{av numb for limits}
	\end{align}
	where
	\begin{equation} \label{Phi_N,M}
		\Phi_{N,M} (\lambda) =  \int\limits_{(N-1)\times (N-1)}\!\!\!\!\! d \widetilde{W} e^{-\frac N2 \Tr \widetilde{W}} (\det  \widetilde{W})^{\frac{M-N-1}{2}} \left| \det \left( \widetilde{W} - \lambda I_{N-1} \right) \right|
	\end{equation}
	To find the averaged total number of roots, we evaluate the above for positive and negative $\lambda$ separately.

	To evaluate $\Phi_{N,M} (\lambda)$ for $\lambda>0$ we diagonalize the $(N-1)\times (N-1)$ positive definite matrix $\widetilde{W}$ as $\widetilde{W}=O\,diag\,\left(s_1,\ldots,s_{N-1}\right)O^{-1}$, where $O$ are orthogonal matrices. Such change induces the transformation: $d\widetilde{W} \propto dO \left| \Delta (s_1, \dots, s_N) \right|\prod_i ds_i$, where $dO$ is the associated Haar measure. As the integrand in \eqref{Phi_N,M} depends only on $s_i$ we can
 rewrite
\begin{equation*}
		\Phi_{N,M} (\lambda>0) = A_{N-1} \int\limits_{\mathbb R_{+}^{N-1}} ds_1\dots ds_{N-1} e^{-\frac N2 \sum\limits_{k=1}^{N-1} s_k} \prod\limits_{k=1}^{N-1} s_k^{\frac{M-N-1}{2}} \left|\Delta_{N-1}(s_1,\dots,s_{N-1})\right| \prod\limits_{k=1}^{N-1} |\lambda-s_k|
	\end{equation*}
where $A_N =  \frac {\pi^{\frac{N(N+1)}4}}{N! \prod_{k=1}^{N}\Gamma \left( \frac{k}2 \right)}$.	Further applying the identity
	\begin{equation*}
	 \left|\Delta_{N-1} (s_1,\dots,s_{N-1})\right| \prod_{k=1}^{N-1} |\lambda - s_k|=\int{ \left|\Delta_N (s_1,\dots,s_N)\right|\delta(\lambda - s_N)}ds_{N}
	\end{equation*}
implies the relation
	\begin{align}\label{intermediate}
		\Phi_{N,M} =&  A_{N-1} \frac{e^{\frac N2 \lambda}}{\lambda^{\frac{M-N-1}{2}}}  \int\limits_{\mathbb R_{+}^N} ds_1\dots ds_N e^{-\frac N2 \sum\limits_{k=1}^{N} s_k} \prod\limits_{k=1}^{N} s_k^{\frac{M-N-1}{2}} \left|\Delta_{N}(s_1,\dots,s_N)\right|\delta(\lambda - s_N)
\end{align}
Finally, by  recalling the joint probability density of Wishart eigenvalues for $N\times N$ case:
\begin{equation} \label{eigenWish2}
			\mathcal P(s_1, \dots, s_N) = C^{-1}_{N,M} e^{-N\sum\limits_{l=1}^N \frac{s_l}2} \prod\limits_{l=1}^N s_l^{\frac{M-N-1}2} \left| \Delta (s_1, \dots, s_N) \right|,
		\end{equation}
where
		\begin{equation}\label{C
_tilde}
			C_{N,M} = \left(\frac 2N\right)^{MN/2} \frac{N!}{\pi^{N/2}} \prod\limits_{k=1}^N \Gamma \left( \frac k 2 \right) \Gamma \left( \frac{M-N-2+k}2 \right),
		\end{equation}
we recognize the  integral in \eqref{intermediate} as related to the mean eigenvalue density $\langle\rho_N (\lambda)\rangle$ of the $N\times N$  Wishart ensemble:
	\begin{equation}\label{34}
		\Phi_{N,M} (\lambda>0) =A_{N-1} C_{N,M} \frac{e^{\frac N2 \lambda}}{\lambda^{\frac{M-N-1}{2}}} \frac{\langle\rho_N (\lambda)\rangle}{N}
	\end{equation}
	Collecting together constants  we  arrive to the following representation of the mean number of positive Lagrange multipliers:
	\begin{align}
		\langle \mathcal N \rangle^{^+}_\sigma =& \sqrt{\frac{2N}{\pi \sigma^2}} (1+\sigma^2)^{-\frac{M+N-2}4} \int\limits_{0}^\infty d\lambda \, \sqrt \lambda e^{N\lambda \left(\frac 12 + \frac 1{\sigma^2}\right)} \langle \rho_N(\lambda)\rangle  \notag\\
		&\times \int\limits_0^\infty dq \, q^{\frac{M-N-2}2} e^{-\frac N{2\sigma^2}|\lambda| \sqrt{1+\sigma^2} \left( q+\frac 1q\right)} \label{av N + sigma}
	\end{align}
	Finally, changing the variables as $ q=e^t$ and $\sqrt{1+\sigma^2}=e^{\delta}$
and using the known integral for the Bessel-Macdonald function $K_\alpha (x)$ allows to represent \eqref{av N + sigma} in the form equivalent to \eqref{main}:
	\begin{align}
		\langle \mathcal N \rangle^{^+}_\delta =&  \sqrt{\frac{N}{\pi }} \frac{e^{-\frac{M+N-1}2 \delta}}{\sqrt{\sinh{\delta}}}  \int\limits_{0}^\infty d\lambda \, \sqrt \lambda e^{\frac{N\lambda}2 \coth{\delta}} \langle \rho_N(\lambda)\rangle   \int\limits_{-\infty}^\infty dt \,  e^{\frac{M-N}2 t-\frac {N \lambda }{2\sinh{\delta}} \cosh{t} } \label{N+ delta}\\
		=&2 \sqrt{\frac{N}{\pi}} \frac{e^{-\frac{M+N-1}{2}\delta}}{\sqrt{\sinh{\delta}}} \int_0^{\infty} d\lambda \sqrt{\lambda} \left\langle\rho_N(\lambda)\right\rangle e^{\frac{N\lambda}{2} \coth \delta } K_{\frac \nu 2} \!\left( \frac{N \lambda}{2 \sinh{\delta}}\right) \label{N+ delta Bessel}
	\end{align}
where we denoted $\nu = M-N$.	
Note that for any given $N,M$ one can most readily evaluate the above integral numerically by using 	
an explicit representations for the Wishart mean eigenvalue density in terms of Laguerre polynomials, see e.g. \cite{LNVbook}.

	Now we proceed to evaluating $\Phi_{N,M} (\lambda)$ from \eqref{Phi_N,M} for $\lambda<0$ where it can not be
related to the mean Wishart eigenvalue density $ \left\langle\rho_N(\lambda)\right\rangle$ as such density is vanishing for negative $\lambda$.  Passing to spectral decomposition and JPD of Wishart eigenvalues in the same way as in the derivation of \eqref{34}, and making change of variables $s_k \rightarrow {x_k}/{N}$ we can write
\begin{align}
		\Phi_{N,M}(\lambda<0) =& \frac{A_{N-1}}{N^{(N-1)\left( 2+\frac{M-N-1}{2} + \frac{N-2}{2} \right)
		}} \int\limits_{\mathbb R_{+}^{N-1}} dx_1\dots dx_{N-1}  \prod\limits_{k=1}^{N-1} x_k^{\frac{M-N-1}{2}} e^{-\frac 12 x_k} \nonumber \\
		\times \left|\Delta_{N-1}(x_1,\dots,x_{N-1})\right| & \prod\limits_{k=1}^{N-1} (x_k+N|\lambda|)
		= \frac{A_{N-1} \widetilde C_{N-1,M-1}}{N^{\frac{(N-1)(M+1)}2}} \left\langle  \det \left[ N|\lambda| I_{N-1} + X^TX \right]  \right\rangle_{X \, wish}\label{negative}
	\end{align}
	where $ I_{N-1}$ is the identity matrix, and $X$ is an $(M-1) \times (N-1)$ matrix with independent, identically  distributed Gaussian real entries, with mean zero and variance equal to one, hence $X^T X$ is the standard Wishart
in an appropriate normalization. Next we use the relation of the determinant in \eqref{negative} to the so-called ''chiral determinant'' form and its relation to the Laguerre polynomials, see \cite{ForrChir}:
	\begin{align*}
		\left\langle  \det \left[ \sqrt{N |\lambda|} I_{N-1} - X^TX \right]  \right\rangle =& (N |\lambda|)^{\frac{N-M}2} \left\langle \det \begin{pmatrix}
			\sqrt{N |\lambda|} I_{N} & X^T \\
			X & \sqrt{N |\lambda|} I_{M}
		\end{pmatrix} \right\rangle \\
		=&  (-1)^N N! L^{M-N}_N (N |\lambda|)
	\end{align*}
	 where generalized Laguerre polynomials are defined via
	\begin{equation}
		L^{M-N}_N (x) = \sum\limits_{k=0}^N \binom{M}{N-k} \frac {(-x)^k}{k!}
	\end{equation}
	Substituting these relations back to \eqref{av numb for limits} we arrive at:
	\begin{align}
		\langle \mathcal N \rangle^{^-}_\sigma =& \frac{ N!N^{\frac{M-N+2}{2}}  }{2^{\frac{M+N-2}{2}}}  \frac{1}{\Gamma\left(\frac N 2\right)\Gamma \left( \frac{M}2 \right)} \frac{1}{\sigma} \frac 1{(1+\sigma^2)^{\frac{M+N-2}{4}}}  \notag\\
		&\times \int\limits_{0}^\infty d\lambda\, e^{-\frac{N\lambda}{\sigma^2}} \lambda^{\frac{M-N}2} \sum\limits_{k=0}^{N-1} \binom{M-1}{N-1-k} \frac{(N \lambda)^k}{k!} K_{\frac \nu 2} \left( N \frac{\sqrt{1+\sigma^2}}{\sigma^2} \lambda \right) \label{N- sigma}
	\end{align}
	The integral over $\lambda$ in \eqref{N- sigma} can be further expressed via the hypergeometric function using the identity \cite{GR}
	\begin{equation}
		\int\limits_0^\infty e^{-\alpha x} K_{\nu}(\beta x) x^{\mu-1} dx = \frac{\sqrt{\pi}(2\beta)^\nu}{(\alpha+\beta)^{\mu+\nu}} \frac{\Gamma(\mu+\nu)\Gamma(\mu-\nu)}{\Gamma(\mu+\frac 12)} F\! \left( \mu+\nu, \nu+\frac 12,\mu+\frac 12, \frac{\alpha-\beta}{\alpha+\beta} \right)
	\end{equation}
	In our case the parameters are $\alpha = \frac N{\sigma^2}$, $\beta = N \frac{\sqrt{1+\sigma^2}}{\sigma^2}$, $\nu = \frac{M-N}{2}$ and $\mu = \frac{M-N}{2}+k+1$, implying that \eqref{N- sigma} can be expressed as shown in
\eqref{subleading}.

 Recall that the $\langle \mathcal N \rangle^{^-}_\sigma$ is the number of \textit{negative solutions} of $g(\lambda) = \frac 1{\sigma^2}$, where from \eqref{equlambda_new_xi_1}
	\begin{equation}
		g(\lambda) = \frac 1N \sum\limits_{i=1}^N \frac{s_i (\bm \xi^T \bm v_i)^2}{(\lambda-s_i)^2}
	\end{equation}
	From the figure \ref{fig g for N=5} it is clear that in every realization
	\begin{equation}
		\mathcal N_\sigma^{^-} = \begin{cases}
			0, &\mbox{ if } \sigma^2 < 1/g(0),\\
			1, &\mbox{ if } \sigma^2 > 1/g(0),
		\end{cases}
	\end{equation}
	where
	\begin{equation}
		g(0) = \frac 1N \sum\limits_{i=1}^N \frac{(\bm \xi^T \bm v_i)^2}{s_i}
	\end{equation}
	One can verify that $g(0)$  is ``self-averaging'' as $N \to \infty$, i.e. tends to a non-random limit $\bar g(0))$ conciding with its mean value. Using that components of $\bm \xi$ are i.i.d. Gaussian mean-zero unit variance variables and $\bm v_\alpha$ are normalized implies $\left\langle(\bm \xi^T \bm v_i)^2\right\rangle =\sum\limits_{\alpha } \bm v_\alpha^2 = 1$, and as they are independent of $s_i$ we have
	\begin{align*}
		 \bar g(0) =  \lim_{N\to \infty} \left\langle \frac 1N \sum\limits_{i=1}^N \frac 1{s_i} \right\rangle =   \lim_{N\to \infty} \frac 1N \int  \left\langle\rho_N(\lambda)\right\rangle\,\frac{d\lambda}{\lambda} = \frac{1}{\alpha-1}
	\end{align*}
where we used the Marchenko-Pastur density \eqref{MP} for explicit evaluation of the integral (cf. Appendix, \eqref{Gres}).  This implies the mean number of negative Lagrangian multiplier is asymptotically given by \eqref{47}. As to the asymptotic value for 	the number of positive Lagrange multipliers, it depends
crucially on the chosen scaling for the noise parameter $\sigma^2$.
	
	\subsubsection{Bulk Marchenko-Pastur regime.}\label{Bulk regime, Marchenko-Pastur}
	
	We start by rewriting the expression \eqref{N+ delta} for $\langle \mathcal N \rangle^{^+}_\delta$ in a form convenient for asymptotic analysis:
	\begin{align}
		\langle \mathcal N \rangle^{^+}_\delta
		=&   \sqrt{\frac{N}{\pi\sinh{\delta}}} e^{-\frac{M+N-1}{2}\delta} \int\limits_0^\infty d\lambda \, \lambda^{\frac 12} \langle \rho_N (\lambda) \rangle e^{\frac{N\lambda}2 \coth \delta} \mathcal J_N(\lambda), \label{N+ delta before laplace n.1}\\
		&\mbox{where } \mathcal J_N(\lambda) = \int\limits_{-\infty}^{\infty} dt \exp\left\{-\frac{N\lambda}2 \frac{\cosh t}{\sinh \delta} + \frac{M-N}2 t\right\}
	\end{align}
	Approximating  the integral $\mathcal J_N(\lambda)$ over $t$ using the Laplace/saddle point method (see Appendix for a derivation) gives
	\begin{equation}\label{mathcal J saddle point}
		\mathcal J_{N>>1}(\lambda) = \frac{2 \sqrt{\pi \sinh \delta}}{N^{\frac 12}(\lambda^2+\kappa^2)^{\frac 14}} \exp\left\{-\frac N2 (\alpha-1)\left(\frac{\sqrt{\lambda^2 + \kappa^2}}{\kappa} - \log
			\left[ \frac{\kappa+ \sqrt{\lambda^2+\kappa^2}}{\lambda} \right]\right)\right\},
	\end{equation}
	where we defined $\kappa = (\alpha-1)\sinh \delta$. Substituting this back to \eqref{N+ delta before laplace n.1} yields the leading asymptotic as $N\gg 1$ in the form:
	\begin{align}\label{N+ delta after laplace n.1A}
		\langle \mathcal N \rangle^{^+}_\delta \approx& 2 e^{-\frac{N(\alpha+1)\delta}{2}} \int\limits_0^\infty d\lambda \, \sqrt{\frac{\lambda}{\sqrt{\lambda^2+\kappa^2}}}\, \langle \rho_N (\lambda) \rangle \, \exp\left\{-\frac{N(\alpha-1)}{2}\mathcal L(\lambda)\right\}
\end{align}
where
\begin{align}
\mathcal L(\lambda)=   \frac{\sqrt{\lambda^2+\kappa^2}}{\kappa} - \log\left[\frac{\kappa + \sqrt{\lambda^2+\kappa^2}}{\lambda}\right] - \lambda \frac{\sqrt{(\alpha-1)^2 + \kappa^2}}{(\alpha-1)\kappa}  \label{N+ delta after laplace n.1}
	\end{align}
This form suggests to use the Laplace method again in \eqref{N+ delta after laplace n.1A}. Recalling that for $N\to \infty$ and fixed $\alpha = \frac MN$ the Marchenko-Pastur distribution \eqref{MP} is nonzero only for $\lambda\in [s_-,s_+]$, with $s_\pm = \left(\sqrt{\alpha} \pm 1\right)^2$, we first look for the stationary point  $\lambda_\ast \in [s_-,s_+]$ of $\mathcal L(\lambda)$. One indeed finds $\lambda_*=\alpha-1$, however the second derivative $\frac{d^2\mathcal L}{d\lambda^2}\bigr|_{\lambda_\ast}$ turns out to be negative, see the Appendix, implying the stationary point is actually a maximum of  $\mathcal L(\lambda)$. Hence, we expect that for finite $\kappa = o(1)$ the integral in \eqref{N+ delta after laplace n.1} will be dominated by regions outside the support of the Marchenko-Pastur distribution.  We however find that this is not the most interesting regime, as it will correspond to the
 complete ''landscape trivialization'', with only two stationary points of the cost function on the sphere.

 We are instead mostly interested in the  scaling  $\delta\rightarrow 0$  for $N\rightarrow\infty$ keeping $\gamma = \frac{\delta N}{4}<\infty$ finite. In such a case $\kappa = o\left(\frac 1N \right)<< 1$ and we can approximate $\left\langle \mathcal N \right\rangle_\delta^{^+}$ from \eqref{N+ delta after laplace n.1} by its leading order contribution using that in such a regime (see Appendix)
	\begin{equation}\label{58}
		e^{-\frac N2 (\alpha-1) \mathcal L (\lambda)} \approx   e^{\gamma  \left[ \frac{(\alpha -1) ^2}{\lambda}  +\lambda \right]}
	\end{equation}
which after a simple rearrangement leads to \eqref{mainlim}. From that formula we can see that in the regime $\delta \sim 1/N$ the number of stationary points in the cost landscape is of order $N$ and decreases when the parameter $\gamma$ (hence $\delta$) increases. In particular, in the limit $\gamma >> 1$  the integral in \eqref{mainlim} is dominated by the vicinity of edges $\lambda=s_-$ and $\lambda=s_+$. To compute the leading order contribution there we scale $\lambda = s_\pm \mp \frac u\gamma$ and find the total contribution to be given by
\eqref{mainlimasy}, see Appendix.  Though we assumed $\gamma$ to be fixed as $N\to \infty$ we may informally consider \eqref{mainlimasy} for the values  $\gamma^{3/2}\sim N$ ( which gives $\delta \sim N^{-1/3}>>N^{-1}$) and conclude that the mean number $\langle \mathcal N \rangle^{^+}$ of
stationary points in the landscape for such a parameter region should be of order unity rather than $N$. This defines another scaling regime where we already can not rely upon the Marchenko-Pastur formula, but should take care of a small vicinity of the edges $s_{\pm}$, which we do next.
	\subsubsection{Edge scaling regime}\label{Edge Scaling}
	The ``edge scaling'' describes the region around spectral edges $s_\pm$ so that the eigenvalues are in a distance of a few level spacings from them. Assuming the parameter $\delta \sim N^{-1/3}$, one can show that one needs the expressions for the mean eigenvalue density for such $\lambda$ where
	\begin{equation} \label{edges}
		\begin{cases}
			\lambda = s_+ + N^{-\frac 23}\left(\frac{4s_+^{2}}{(s_+-s_-)}\right)^{\frac 13} \xi \\
			\lambda = s_- - N^{-\frac 23}\left(\frac{4s_-^{2}}{(s_+-s_-)}\right)^{\frac 13}  \xi
		\end{cases}
	\end{equation}
	The eigenvalue density is then given by the expressions which can be found in equations (3.31)-(3.32) of the paper \cite{ForrLD}:
	\begin{equation}
		\left\langle \rho (\lambda)\right\rangle \rightarrow \left(\frac{s_+-s_-}{4Ns_\pm^{2}}\right)^{\frac 13} \rho_{edge} (\xi)
	\end{equation}
where the density $\rho_{edge} (\xi)$ was given in \eqref{edgedens} in terms of the Airy functions.

	To analyse the expression \eqref{N+ delta after laplace n.1A} for the average number of positive Lagrange multipliers $ \left\langle \mathcal N \right\rangle^{^+}_\delta$   we need to look for the higher-order contributions from around the upper edge $s_+$ and the lower edge $s_-$ and combine those. Reparametrizing our noise parameter as $\omega= N^{\frac 13} \delta \left( \dfrac{s_+-s_-}{4}  \right)^{\frac 23} \in [0,\infty]$ and keeping $\omega$ finite as $N\to \infty$ we arrive at (see Appendix):
	\begin{equation}\label{av num edge sum}
		\lim_{N\rightarrow \infty } \left\langle \mathcal N \right\rangle^{^+}_\omega = 2 \int\limits_{-\infty}^{\infty} \left[ \exp\left\{-\frac{\omega^3}{3s_-} + \frac{\omega\xi}{s_-^{1/3}}\right\} +\exp\left\{-\frac{\omega^3}{3s_+} + \frac{\omega\xi}{s_+^{1/3}}\right\} \right] \rho_{edge} (\xi) d \xi
	\end{equation}
	In \cite{FLD2013} it has been shown that asymptotically
	\begin{equation}
		e^{-\frac{x^3}{24}} \int\limits_{-\infty}^\infty e^{\frac x2 \xi } \rho_{edge} (\xi) d\xi \approx
		\begin{cases}
			\sqrt{\frac 2\pi} {x^{-\frac 32}}, &\mbox{ for } x <<1\\
			\frac 12, &\mbox{ for } x>>1
		\end{cases}
	\end{equation}
	Applying this to our case, with $x_{\pm} = 2\omega s_{\pm}^{-1/3}$, we see that for $\omega << 1$
	\begin{equation}\label{65}
		\left\langle \mathcal N \right\rangle^{^+}_\omega \approx N \frac{1}{2 \sqrt\pi \gamma^{3/2} }
	\end{equation}
	This agrees perfectly with our findings for the bulk regime with $\gamma>>1$ in \eqref{mainlimasy}. On the other hand, for $\omega >> 1$, we have $ \left\langle \mathcal N \right\rangle^{^+}_\omega \rightarrow 2$, implying that for any fixed and finite variance $\sigma^2$ there are only two stationary points, one of the Lagrange multipliers corresponding to a minimum and the second one to the maximum.
	
	\subsection{Large deviations for the minimal Lagrange multiplier}\label{Large deviations}
We start with recalling the asymptotic density $p_{+}(\lambda)$ of positive Lagrange multipliers for $N\gg 1$ implied by \eqref{N+ delta after laplace n.1A}:
 \begin{align}
	p_{+}(\lambda)	 \approx& 2 e^{-\frac{N(\alpha+1)\delta}{2}} \sqrt{\frac{\lambda}{\sqrt{\lambda^2+\kappa^2}}}\, \langle \rho_N (\lambda) \rangle \, \exp{-\frac{N}{2}\mathcal L_1(\lambda)}
\label{p+ delta}
	\end{align}
where $\left\langle \rho(\lambda)\right\rangle$ is the mean density of Wishart eigenvalues  and
\begin{equation}\label{L1}
L_1 (\lambda) = (\alpha - 1)\left[ \frac{\sqrt{ \lambda^2+\kappa^2}}{\kappa} - \log \left( \kappa + \sqrt{ \lambda^2 + \kappa^2} \right) -  \lambda \frac{\sqrt{(\alpha - 1)^2 + \kappa^2}}{(\alpha -1 ) \kappa}   \right]
\end{equation}

 We know that the lowest Lagrange multiplier $\lambda_1:=\lambda_{min}$ is the only one located to the left of the smallest eigenvalue of the Wishart matrix. In the limit $N\gg 1$ the Wishart eigenvalues concentrate in the spectral interval $[s_-,s_+]$, hence all other Lagrange  multipliers but $\lambda_{min}$ with overwhelming probability belong to the same interval $[s_-,s_+]$ as well.  It is therefore clear that for $N\gg 1$ the Lagrange multiplier density $p_+(\lambda)$ for $0<\lambda<s_{-}$ should asymptotically be the same as the probability density of $\lambda_{min}$, so we identify the two densities below.

The mean density of Wishart eigenvalues  for large but finite $N$ outside of the support of the Marchenko-Pastur density \eqref{MP} is known to take a Large Deviation form $\langle \rho_N (\lambda) \rangle \propto e^{-\frac N2 L_2 ( \lambda)}$, see eq.(1.4) in \cite{ForrLD}, where
\begin{align}
		L_2 (\lambda) =& - \sqrt{(\lambda - s_-)(\lambda - s_+)} - 2\log (\alpha + 1 - \lambda + \sqrt{(\lambda - s_-)(\lambda - s_+)}) \notag \\
		&+ 2 (\alpha -1) \log(\lambda + \alpha -1 + \sqrt{(\lambda - s_-)(\lambda - s_+)}). \label{L_2}
	\end{align}
Based on \eqref{p+ delta} we then conclude that for $0<\lambda<s_{-}$ the Lagrange multiplier density $p_+(\lambda)$ has the Large deviation form
	\begin{equation}\label{LD func}
		p_+( \lambda) \propto e^{-\frac N2 \Phi ( \lambda)},
	\end{equation}
	where $\Phi (\lambda) = L_1 (\lambda) + L_2 (\lambda) + const$ and
\[
const= \frac{\alpha +1}{2} \log(1+\sigma^2) + 2(\alpha - 2) \log{\frac{1}{2\sqrt{\alpha}}}\,.
\]		
	We would like to find such $\lambda=\lambda_*$ that minimises $\Phi (\lambda)$, which should then
provide us with the most probable/typical value of the smallest Lagrange multiplier.
For this we need to solve $\frac{d}{d\lambda}\Phi (\lambda)=0$, which turns out to be hard when
attempted directly. We found a bypass for this difficulty by determining $\lambda_*$ in an
alternative approach, and then simply verify that indeed such value minimizes  $\Phi (\lambda)$.

 Recall that we are after the typical value $\lambda_\ast$ of the smallest solution of the equation
 $g(\lambda) = \frac{1}{\sigma^2}$ where from \eqref{equlambda_new_xi_1}
	\begin{equation}\label{g}
		g(\lambda) = \frac 1N \sum\limits_{i=1}^N \frac{s_i (\bm \xi^T \bm v_i)^2}{(\lambda-s_i)^2}.
	\end{equation}
The eigenvectors $\bm v_i$ and eigenvalues $s_i$ being random,  $g(\lambda)$ is random
as well. However we know that the smallest solution $\lambda_{min}$ is located {\it outside} the spectrum of the Wishart matrix, and for such $\lambda$ the matrix elements of the resolvents like $g(\lambda)$ are well-known to be non-fluctuating (selfaveraging) as $N\to \infty$. We therefore conclude that to find $\lambda_\ast$ as $N\to \infty$ should be simply equivalent to solving the ensemble-averaged equation
\begin{equation}
		\left\langle\left\langle g(\lambda_\ast)\right\rangle_{\bm b}\right\rangle_{W} = \frac{1}{\sigma^2}, \quad \lambda<s_{-}.
	\end{equation}
After some algebraic manipulations presented in the Appendix  the averaged equation in the range $\lambda<s_{-}$ takes the following form
	\begin{equation}\label{49}
		\left\langle\left\langle g(\lambda)\right\rangle_{\bm b}\right\rangle_{W}=  \frac{1}{(\sqrt{s_+}-\sqrt{s_-})^2}
		\frac{\left(\sqrt{s_+-\lambda} - \sqrt{s_- - \lambda }\right)^2}{\sqrt{(s_+-\lambda)( s_- - \lambda )}} = \frac 1{\sigma^2}
	\end{equation}
	Solving this in a straightforward way one finds
	\begin{equation}\label{lambda ast}
		\lambda_\ast = (\alpha +1) - \sqrt{\alpha} \left( \sqrt{1+\sigma^2} +  \frac{1}{\sqrt{1+\sigma^2}} \right) = \frac{\left(\sqrt \alpha - \sqrt{1+\sigma^2}\right) \left( \sqrt{\alpha (1+\sigma^2)}-1  \right)}{\sqrt{1+\sigma^2}}.
	\end{equation}
	 As a consistency check, we see that for $\sigma^2 \rightarrow 0$ one has $\lambda_\ast \rightarrow \alpha+1 - 2\sqrt \alpha = (\sqrt \alpha - 1)^2=s_- $ as expected, since in this limit minimal $\lambda$ should coincide with the minimal eigenvalue of the Wishart matrix. Also $\lambda_\ast =0$ for $\sigma^2 = \alpha - 1$
agreeing with our conclusions in \eqref{47}.

	Next, we check that the Large Deviation rate $Phi(\lambda)$ is indeed stationary at the value $\lambda_*$, that is  $\frac{d\Phi}{d\lambda} = 0$ at $\lambda = \lambda_\ast$.  To this end we evaluate $  \frac{dL_1}{d\lambda}\bigr|_{\lambda_\ast} $ and $ \frac{dL_1}{d\lambda}\bigr|_{\lambda_\ast}$ in the Appendix and find
	\begin{equation}\label{eq to 0}
		\left( \frac{dL_1}{d\lambda} +\frac{dL_2}{d\lambda} \right)\Biggr|_{\lambda_\ast} = \frac 1{\lambda_\ast} \left[ (\alpha - 1)\left[\frac{\sqrt{\kappa^2+\lambda^2_\ast}}{\kappa } -1  \right] - \frac{2}{\sigma^2} \left(\sqrt{ 1+\sigma^2}-\sqrt\alpha\right)^2 \right]
	\end{equation}
	After straightforward manipulations we verify in the Appendix that the right-hand side is equal to zero, confirming that $\lambda_\ast$ provides the stationary point (actually, the minimum) of the large deviations function $\Phi (\lambda)$.  Hence $\lambda_\ast$ is the most probable value of the smallest Lagrange multiplier $\lambda_{min}$ and  deviations of $\lambda_{min}$ from this value for large $N\gg 1$  are punished exponentially.

	To conclude this section, we compute the value of the cost/loss function at $\lambda_\ast$, which (assuming self-averaging again) should give us the typical minimal cost.  We start with substituting our expression for the position of a stationary point $\bm x = (W-\lambda I)^{-1}A^T \bm b$ to the cost \eqref{energy} and get after simple manipulations:
	\begin{equation}\label{52}
		\frac{\mathcal E_{\lambda}}{N}= \frac 12 \left[ \lambda +\frac{1}N \bm b^2 +  \frac 1N \bm b^TA \frac 1{\lambda I-W} A^T \bm b \right]
	\end{equation}
We calculate the mean value $\langle \mathcal E_{\lambda} \rangle / N$ at $\lambda_*$ by first averaging over the normally-distributed vector $\bm b$ so that ${\langle \bm b^T R \bm b \rangle_{ \bm b} = \sigma^2 \Tr R}$:
	\begin{align*}
		\frac{\langle \mathcal E_{\lambda_\ast} \rangle}{N} =& \frac 12 \left[ \lambda_\ast +\alpha \sigma^2 + \sigma^2\frac 1N  \left\langle \Tr{\frac{W}{\lambda_* - W}} \right\rangle_{\!\!\!wish} \right]
		=\frac 12 \left[ \lambda_* + \lambda_* \sigma^2\frac 1N  \left\langle \Tr{\frac{1}{\lambda_* - W}} \right\rangle + (\alpha-1)\sigma^2 \right]
\end{align*}
and then may use the known expression for the mean resolvent trace of the Wishart matrix as $N\to \infty$:
\begin{align*}
	\lim_{N\to \infty}\frac{\langle \mathcal E_{\lambda_\ast} \rangle}{N}=	& \frac 12 \left[ \lambda_* + \sigma^2\frac 2{(\sqrt{s_+}-\sqrt{s_-})^2}  \left[ \lambda_* + \sqrt{(\lambda_* - s_-)(\lambda_* - s_+)} - \sqrt{s_+s_-} \right] +(\alpha-1)\sigma^2\right]\\
		=&\frac 12 \left[ \lambda_* + \frac{\sigma^2}2  \left[ \lambda_* + \sqrt{(\lambda_* - s_-)(\lambda_* - s_+)} - (\alpha- 1) \right] +(\alpha-1)\sigma^2\right].
	\end{align*}
	The expression \eqref{lambda ast} for $\lambda_*$  implies that
	\begin{equation*}
		\sqrt{(\lambda_* - s_-)(\lambda_* - s_+)} = \alpha+1-\frac{2\sigma^2}{\sqrt{1+\sigma^2}}
	\end{equation*}
 In this way we find after straightforward algebra that the mean  value of the minimal loss function is indeed given by \eqref{minmeanvalue}. We will confirm this result by independent calculation in the next section, and recover
the Large Deviation rate describing the probability of deviations of the minimal cost from its typical value. In this way we will see that the above is not only the mean but simultaneously equal to the typical/most probable value of the minimal cost.

\subsection{Derivation of the minimal cost and its Large Deviation function by replica method} \label{Large deviations for the minimal cost}

In the introduction we described the ideas of the method which is based
on the powerful albeit heuristic method of Theoretical Physics, known as the ''replica trick''.
  In that framework one evaluates the ''replicated'' disorder averaged partition function $\langle Z^n  \rangle$  and subsequently taking the limit $n\to 0$ recovers the averaged log, see \eqref{replica method_intro}.
The details of evaluating $\langle Z^n  \rangle$ in a closed-form will be postponed to
 the Appendix \ref{partion_fun_evaluation} where this is done for a broad class of  random cost/loss functions, for which the cost \eqref{energy} studied in this paper is only the special simplest case. Here we just give the right-hand side represented as:
	\begin{equation}\label{moments <Z^n>}
		\langle Z^n \rangle =C_{N,n} \int\limits_{D_N^{(Q)}} d Q \, (\det Q)^{-\frac{n+1}{2}} e^{  -\frac N2 \Phi_n (Q) } ,
	\end{equation}
	where $Q$ is the matrix with entries originally defined as scaled scalar products $q_{ab} = \frac 1N (\bm x_a \cdot \bm x_b) $ and is, therefore, non-negative definite, and explicit value of the constant $C_{N,n}$ is known but not important at the moment. The domain of integration is then over such matrices $Q$ with the constraint on diagonal entries,  that is given by
\begin{equation}
		Q = \begin{pmatrix}
			1 & & q_{ab} \\
			& \ddots & \\
			q_{ab} & & 1
		\end{pmatrix}.
	\end{equation}
In short, the integration goes over the domain   $D_N^{(Q)} = {\{ Q\geq 0, q_{aa}=1 \, \forall a\}}$. The function in the exponent is
\begin{equation}\label{special case}
\Phi_n (Q) \equiv \alpha \log \det \{ I_n + \beta (Q+\sigma^2 E_n)\} - \log \det Q
\end{equation}
 where ${\alpha=M/N}$ and $E_n$ stands for the $n\times n$ matrix with all entries equal to unity, whereas
 $I_n$ stands for the $n\times n$ identity matrix.\\
Due to the presence of large factor $N$ in the exponent under the integral over $Q$ in \eqref{moments <Z^n>}, we can apply the Laplace's method approximation  to the integral over $Q$, which then implies that for some minimizing matrix argument $Q_{min}$
\begin{equation}\label{leading Z^n}
	\langle Z^n \rangle \propto e^{-\frac N2 \Phi_n (Q_{min})}
\end{equation}
where we keep only the leading exponential terms, as only them are needed for finding $ \langle  \mathcal E_{min} \rangle$ to the leading order:
	\begin{equation}\label{mean_minimum}
	\lim_{N\to \infty}	\frac{\langle \mathcal E_{min} \rangle}N =\lim_{\beta \to \infty} \frac{1}{2\beta} \lim\limits_{n\rightarrow 0} \frac{1}{n} \Phi_n (Q_{min})
	\end{equation}
  Note that one can argue in general that for a large class of random cost functions the typical and the mean values for the minimum coincide (this fact is frequently referred to as the ``free energy self-averaging'' property).\\
To search for $Q_{min}$ we use the stationarity conditions: $\frac{\partial  \Phi_n (Q)}{\partial Q_{ab}}=0, \forall a<b$ which gives the equation:
\begin{equation}\label{stationarity_simple}
\left(Q^{-1}\right)_{ab}=\alpha \beta \left(I_n + \beta (Q+\sigma^2 E_n)\right)^{-1}
\end{equation}
To solve this equation we use the so-called {\it Replica Symmetric Ansatz} (see a discussion about its validity in the present context in the end of Conclusion and Perspectives section) which amounts to assuming that all diagonal entries of the matrix $Q$ are equal to one and all other entries are same number $0\le q<1$. For such a matrix the inverse is well-known:
\begin{equation}\label{Qinverse}
\left(Q^{-1}\right)_{aa}=\frac{\left(1+q(n-1)\right)}{(1-q)\left(1+q(n-1)\right)}, \quad \left(Q^{-1}\right)_{a<b}=-\frac{q}{(1-q)\left(1+q(n-1)\right)},
\end{equation}
as well as the determinant
\begin{equation}\label{detQ}
\det Q=\left(1+q(n-1)\right)(1-q)^{n-1}
\end{equation}
Similarly, it is easy to find the inverse of the matrix $R=I_n + \beta (Q+\sigma^2 E_n)$ since this matrix has
diagonal entries all equal to $R_{aa}=1+\beta(1+\sigma^2)=:r_d,\, \forall a=1,\ldots,n$ and all off-diagonal entries  equal to $R_{ab}=\beta(q+\sigma^2)=:r, \, \forall a<b$. The inverse of such matrices is given by
\begin{equation}\label{Rinverse}
\left(R^{-1}\right)_{aa}=\frac{\left(r_d+r(n-1)\right)}{(r_d-r)\left(r_d+r(n-1)\right)}, \quad \left(R^{-1}\right)_{a<b}=-\frac{r}{(r_d-r)\left(r_d+r(n-1)\right)},
\end{equation}
Hence the stationarity condition takes the form:
\begin{equation}\label{stationarity_RS_simple}
\frac{q}{(1-q)\left(1+q(n-1)\right)}=\alpha \frac{\beta^2(q+\sigma^2)}{(1+\beta-\beta q)(1+\beta-\beta q+n\beta(q+\sigma^2)}
\end{equation}
The corresponding value of $\Phi_n (Q)$  within the replica symmetric ansatz using \eqref{detQ} can be easily written as
 \begin{equation}\label{Phi_RSa}
\Phi_n (Q)=\alpha n\log{\left(1+\beta(1-q)\right)}+\alpha\log{\left(1+\frac{\beta(1+\sigma^2)n}{1+\beta(1-q)}\right)}
-n\log{(1-q)}-\log{\left(1+q\frac{n}{1-q}\right)}
\end{equation}
So far this consideration was exact for any positive integer $n$. Now we need to use that we are seeking to perform in the end the replica limit $n\to 0$. In this limit
\begin{equation}\label{Phi_RSb}
\lim_{n\to 0}\frac{1}{n}\Phi_n (Q)=\alpha \log{\left(1+\beta(1-q)\right)}+\frac{\beta(1+\sigma^2)}{1+\beta(1-q)}-\log{(1-q)}-\frac{q}{1-q}
\end{equation}
We also can seek for the solution of \eqref{stationarity_RS_simple}
directly in the replica limit setting $n=0$ in the equation, giving:
\begin{equation}\label{stationarity_RS_simple_replica}
\frac{q}{(1-q)^2}=\alpha\frac{\beta^2(q+\sigma^2)}{(1+\beta-\beta q)^2}
\end{equation}
For finite $\beta$ the equation is equivalent to a cubic one. To simplify our consideration further we recall that to find the minimum of the cost function from \eqref{mean_minimum} we only need to know $q$ in the limit $\beta \to \infty$. Two different situations may happen in this limit
\begin{enumerate}
\item  $q$ tends in the limit $\beta\to \infty$ to a non-negative value smaller than unity
(this range is dictated by positivity of the matrix $Q$).
\item  Alternatively, in such a limit $q$ tends to unity in such a way that
$v=\lim_{\beta \to \infty} \beta (1-q)$ remains finite.
 \end{enumerate}
 One may notice that taking the limit $\beta\to \infty$  via using \eqref{mean_minimum} in the first case (when the limiting $q$ is smaller than unity)
yields  $\lim_{N\to \infty}	\frac{\langle \mathcal E_{min} \rangle}N=0$. To understand for which
values of parameters this situation take place we perform the required limit $\beta\to \infty$ in \eqref{stationarity_RS_simple_replica}. This immediately
 produces the relation $q=\alpha(q+\sigma^2)$ with the solution $q=\frac{\alpha\sigma^2}{1-\alpha}$.
 The inequality $0\le q<1$ gives the condition $0<\alpha<\alpha_c:=1/(1+\sigma^2)<1$. Thus for such range of
 the ratio $\alpha=M/N$ the system of linear equations on the sphere is typically compatible, i.e. has a solution, and hence the minimal cost is vanishing. For $\alpha>\alpha_c$ we expect that $v=\lim_{\beta \to \infty} \beta (1-q)$ remains finite and we will see that in such a situation the minimum cost is positive, signalling of incompatibility of equations.

Instead of evaluating the minimal cost for $\alpha>\alpha_c$ we directly address the calculation of the whole Large Deviation function. The corresponding formalism was presented in the Introduction, see in particular
\eqref{LD1}-\eqref{LD_Legendre1}. It is clear by comparing
   \eqref{leading Z^n}  with \eqref{LD_Legendre} that  $\phi(s)$ can be found as
 \begin{equation}\label{LD_replica}
\phi(s)=-\frac{1}{2}\lim_{\substack{n=s/\beta\\ \beta\to \infty}}\Phi_n(Q_{min}),
\end{equation}
 hence we need to evaluate the function $\Phi_n(Q_{min})$
in the limit $n\to 0$ and $\beta\to \infty$ keeping $n\beta=s$ fixed. The starting point of this procedure is the expression \eqref{Phi_RSa} for $\Phi_n (Q)$  as well as the corresponding stationarity condition \eqref{stationarity_RS_simple}. From the previous analysis for $\alpha>\alpha_c$ we expect to have $q\to 1$ as $\beta\to \infty$ in such a way that $v=\beta(1-q)$ remains finite. Correspondingly we substitute $n=s/\beta, q=1-v/\beta$ into \eqref{Phi_RSa} and set $\beta \to \infty$, resulting in the following expression for the  functional:
\begin{equation}\label{Phi_RS_lim}
\Phi (s,v)=  \alpha\log{\left(1+\frac{s(1+\sigma^2)}{1+v}\right)}-\log{\left(1+\frac{s}{v}\right)}
\end{equation}
whereas the stationarity condition  \eqref{stationarity_RS_simple} takes the form
\begin{equation}\label{stationarity_RS_simple2}
\frac{1}{v(v+s)}= \frac{\alpha(1+\sigma^2)}{(1+v)\left(1+v+s(1+\sigma^2)\right)}
\end{equation}
which is in fact equivalent to the condition $\frac{\partial}{\partial v} \Phi (s,v)=0$.
From this point one needs to find $v(s)$ solving  \eqref{stationarity_RS_simple2}
which one can write equivalently as
\begin{equation}\label{stationarity_RS_simple3}
\alpha(1+\sigma^2)v(v+s)= {(1+v)\left(1+v+s(1+\sigma^2)\right)}
\end{equation}
and in this way first get the function $\phi(s)=-\frac{1}{2}\Phi(s,v(s))$. After that one performs the Legendre transform \label{Legendre transform} over the variable $s$ to obtain the large deviation rate via
\begin{equation}\label{LD_rate_def1}
{\cal L}({\bm e })=-{\bm e}s_*+\frac{1}{2}\Phi\left(s_*,v(s_*)\right)=\frac{1}{2}\left[\Phi\left(s_*,v(s_*)\right)-2{\bm e}s_*\right],
\end{equation}
where $s_*$ as a function of ${\bm e}$ is found by solving the equation
\begin{equation}\label{LD_rate_def2}
{\bm e}=\left. \frac{1}{2}\frac{d\Phi}{ds}\right\vert_{s_*}  = \left. \frac{1}{2}\frac{\partial\Phi(s,v)}{\partial s} \right\vert_{s_*},
\end{equation}
where the last equation follows from the above-mentioned stationarity: $\frac{\partial}{\partial v} \Phi (s,v)=0$.
Differentiating \eqref{Phi_RS_lim} we find, using \eqref{stationarity_RS_simple2}, that
\begin{equation}\label{LD_rate_def3}
{\bm e}=\frac{1}{2}\left(\frac{\alpha(1+\sigma^2)}{1+v+s_*(1+\sigma^2)}-\frac{1}{v+s_*}\right)=
\frac{1}{2}\left(\frac{1+v}{v(v+s_*)}-\frac{1}{v+s_*}\right)=\frac{1}{2v(v+s_*)}
\end{equation}
or equivalently
\begin{equation}\label{LD_rate_def4}
v(v+s_*)=\frac{1}{2{\bm e}}
\end{equation}
Using the above one can rewrite
\eqref{stationarity_RS_simple3} as
\begin{equation}\label{stationarity_RS_simple4}
\alpha \left(1+\sigma^2 \right) \frac{1}{2{\bm e}}= {(1+v) \left(1+v+s_*(1+\sigma^2)\right) }
\end{equation}
Further expressing $s_*$ from \eqref{stationarity_RS_simple3} as
\begin{equation}\label{stationarity_RS_simple5}
s_*(1+\sigma^2)=\frac{v^2\left(\alpha(1+\sigma^2)-1\right)-2v-1}{1-v(\alpha-1)}
\end{equation}
and substituting back to \eqref{stationarity_RS_simple4} we get the closed-form equation for $v$ as a function of
${\bm e}$:
\begin{equation}\label{LD_rate_def5}
\alpha(1+\sigma^2)\frac{1}{2{\bm e}}= (1+v)\left\{1+v+\frac{v^2\left(\alpha(1+\sigma^2)-1\right)-2v-1}{1-v(\alpha-1)}\right\}=
\frac{\alpha(1+v)v(v\sigma^2-1)}{1-v(\alpha-1)}
\end{equation}
which can be equivalently rewritten as a cubic equation \eqref{cubic} which we repeat below
\begin{equation}\label{LD_rate_def6}
\sigma^2v^3+v^2(\sigma^2-1)-v\left(1-a(\alpha-1)\right)-a=0, \quad a:=\frac{1+\sigma^2}{2{\bm e}}.
\end{equation}
Solving this equation one gets the value of $v$ for a given ${\bm e}$.
It turns out that the knowledge of $v({\bm e})$ is enough to get the large deviation rate, which can be expressed solely via such $v$. The simplest way to proceed is by taking logarithms of both sides
in the stationarity condition  \eqref{stationarity_RS_simple2} to present it in a form of the identity
\begin{equation}\label{station_ident}
\alpha\log{\left(\frac{1+v+s(1+\sigma^2)}{1+v}\right)}=\alpha  \log{\left(\frac{v(v+s)}{(1+v)^2}\alpha(1+\sigma^2)\right)}
\end{equation}
which allows to rewrite the expression \eqref{Phi_RS_lim} first as
\begin{equation}\label{Phi_RS_lim2}
\Phi (s,v)= \alpha  \log{\left(\frac{v(v+s)}{(1+v)^2}\alpha(1+\sigma^2)\right)}-\log{\left((s+v)v\right)}+2\log{v}
\end{equation}
and then using \eqref{LD_rate_def4} as
\begin{equation}\label{Phi_RS_lim2}
\Phi ({\bm e},v)=-(\alpha-1)\log{(2{\bm e})}+\alpha\log{\left(\alpha(1+\sigma^2)\right)}+2\log v-2\alpha\log{(1+v)}
\end{equation}
This should be combined with the relation
\begin{equation}
{\bm e}s_*=\frac{1}{2v}-{\bm e}v
\end{equation}
following from \eqref{LD_rate_def4}. In this way one completely solves the problem of explicitly inverting the Legendre transform and providing the required  Large Deviation Rate function \eqref{LD mincost_main} for the minimal cost as a function of ${\bm e} $.

\newpage
\section{Appendices}

	This section presents technical detail helping to derive some formulas presented in the main text of the paper.
	
	\subsection{To section \ref{Bulk regime, Marchenko-Pastur}}
\begin{enumerate}
\item 	
	To derive \eqref{mathcal J saddle point} we apply the Laplace method to
	\begin{equation*}
		\mathcal J_N(\lambda) = \int\limits_{-\infty}^{\infty} dt e^{-\frac N2 \mathcal L (t)}, \mbox{ with } \mathcal L(t) = \lambda \frac{\cosh t}{\sinh \delta} - (\alpha - 1) t.
	\end{equation*}
for $N\rightarrow \infty$ assuming $\alpha=\frac M N > 1$ is fixed. Taking the derivative of $\mathcal L (t)$
	\begin{align*}
		\frac{d \mathcal L}{dt} &= \lambda \frac{\sinh{t}}{\sinh{\delta}} - (\alpha -1) = 0
\end{align*}
	we find conditions for the stationary point $t_\ast$ as
\begin{align}
 \sinh{t_\ast} = \frac{(\alpha -1)}{\lambda} \sinh{\delta} = \frac{\kappa}{\lambda}
	\end{align}
where
	 we introduced $\kappa: = (\alpha -1)\sinh{\delta}$ so that $\cosh{t_\ast} = \frac{\sqrt{\lambda^2 + \kappa^2}}{\lambda}$.  The value of second derivative at the stationary point is
	\begin{align*}
		\frac{d^2 \mathcal L}{dt^2}\Bigr|_{t_\ast} &= \lambda \frac{\cosh{t_\ast}}{\sinh{\delta}} = \frac{\sqrt{\lambda^2 + \kappa^2}}{\sinh{\delta}} \geq 0
	\end{align*}
	showing that it is a minimum, as required. The value of the exponent at the stationary point is given by
	\begin{align*}
		\mathcal L (t_\ast) =&  \frac{\sqrt{\lambda^2 + \kappa^2}}{\sinh \delta} - (\alpha - 1) \sinh^{-1}
		\left( \frac \kappa \lambda \right)  \\
		=& (\alpha-1)\left(\frac{\sqrt{\lambda^2 + \kappa^2}}{\kappa} -  \log
		\left[ \frac{\kappa+ \sqrt{\lambda^2+\kappa^2}}{\lambda} \right]\right)
	\end{align*}
	and collecting all factors, we find
	\begin{align*}
		\mathcal J_{N>>1}(\lambda) \approx &e^{-\frac N2 (\alpha-1)\left(\frac{\sqrt{\lambda^2 + \kappa^2}}{\kappa} - \log
			\left[ \frac{\kappa+ \sqrt{\lambda^2+\kappa^2}}{\lambda} \right]\right)} \int\limits_{-\infty}^\infty dt\, e^{-\frac N4 \frac{\sqrt{\lambda^2 + \kappa^2}}{\sinh{\delta}} (t-t_\ast)^2}
	\end{align*}
leading to the announced formula \eqref{mathcal J saddle point}.

	\item Evaluating \eqref{N+ delta after laplace n.1} by Laplace method we compute the derivative
	\begin{align*}
		\frac{d \mathcal L}{d\lambda} =&  \frac{\lambda }{\kappa \sqrt{\lambda^2+\kappa^2}} - \frac{1}{\kappa + \sqrt{\lambda^2+\kappa^2}}\frac{\lambda}{\sqrt{\lambda^2+\kappa^2}} + \frac 1\lambda - \frac{\sqrt{(\alpha-1)^2 + \kappa^2}}{(\alpha-1)\kappa}  \\
		=& \frac{1}{\sqrt{\lambda^2+\kappa^2}} \left(  \frac \lambda\kappa + \frac \kappa\lambda \right) - \frac{\sqrt{(\alpha-1)^2 + \kappa^2}}{(\alpha-1)\kappa}
		=\frac{\sqrt{\lambda^2+\kappa^2}}{\lambda\kappa} - \frac{\sqrt{(\alpha-1)^2 + \kappa^2}}{(\alpha-1)\kappa}
	\end{align*}
	This implies that $\lambda_\ast=(\alpha-1)>0$ is a stationary point, with $\lambda_* \in [s_-, s_+] $. Now we compute the second derivative of $\mathcal L$ at $\lambda_*$:
	\begin{equation*}
		\frac{d^2\mathcal L}{d\lambda^2}\Biggr|_{\lambda_\ast} = \frac{\lambda^2\frac{1}{\sqrt{\lambda^2+\kappa^2}} - \sqrt{\lambda^2+\kappa^2}}{\kappa \lambda^2} = -\frac{\kappa}{\lambda^2\sqrt{\lambda^2+\kappa^2}}<0,
	\end{equation*}
implying that the stationary point is a maximum.
\item 	For \eqref{58} we expand in Taylor series for $\kappa\to 0$:
	\begin{align*}
		\mathcal{L}(\lambda)\bigr|_{\kappa \rightarrow 0} =&  \frac \lambda\kappa \left(1+\frac{\kappa^2}{\lambda^2}\right)^{\frac 12} +\log \lambda - \log\left[\kappa + \lambda\left(1+\frac{\kappa^2}{\lambda^2}\right)^{\frac 12} \right] - \frac\lambda\kappa \left(1 + \frac{\kappa^2}{(\alpha-1)^2}\right)^{\frac 12}  \\
		\approx& \frac \lambda\kappa \left(1+\frac{\kappa^2}{2\lambda^2} + \cdots \right) +\log \lambda - \log\left[\lambda\left(1+\frac\kappa\lambda +\frac{\kappa^2}{2\lambda^2} + \cdots \right) \right] - \frac\lambda\kappa \left(1 + \frac{\kappa^2}{2(\alpha-1)^2} + \cdots \right)  \\
		=& \frac \lambda\kappa+\frac{\kappa}{2\lambda}   +\log \lambda -\log \lambda - \frac\kappa\lambda - \frac{\kappa^2}{2\lambda^2} +\frac{\kappa^2}{2\lambda^2} - \frac\lambda\kappa  - \frac{\lambda\kappa}{2(\alpha-1)^2} + o\!\left(\frac{\kappa^3}{\lambda^3}  \right)  \\
		\approx&-\frac{\kappa}{2\lambda}  - \frac{\lambda\kappa}{2(\alpha-1)^2}
	\end{align*}
\item 	To arrive to \eqref{mainlim} we first have from \eqref{N+ delta after laplace n.1} and \eqref{58}
	\begin{align*}
		\lim_{ N\rightarrow\infty} \frac{\langle \mathcal N \rangle_\delta^{^+}}{N} =& \frac 2N e^{-2\gamma(\alpha+1)} \int\limits_0^\infty d\lambda \,  \langle \rho_N (\lambda) \rangle  e^{\gamma  \left[ \frac{(\alpha -1) ^2}{\lambda}  +\lambda \right]}
\end{align*}
Noticing  that $(\alpha-1)^2=s_+ s_-$ and $2(\alpha+1)=s_++s_-$ and using the Marchenko-Pastur density \eqref{MP} we further have
\begin{align*}
		=&\frac 2N e^{-\gamma (s_+ + s_-)} \int\limits_{s_-}^{s_+} d\lambda \, \frac{N}{2\pi}\,\frac{\sqrt{(\lambda-s_-)(s_{+}-\lambda)}}{\lambda} e^{\gamma  \left[ \frac{s_+ s_-}{\lambda}  +\lambda \right]}
	\end{align*}
equivalent to \eqref{mainlim}.
\item 	To arrive to \eqref{mainlimasy}, we consider constributions from two edges separately.
	In the vicinity of $\lambda=s_-$ we use the scaling $\lambda = s_- + \frac u\gamma$ which allows to
estimate the corresponding contribution as
	\begin{align*}
		&\frac 1\pi \int\limits_{0}^{\infty} \frac{du}{\gamma s_-} \sqrt{\frac u\gamma} \sqrt{s_+-s_-} \, e^{-u\frac{s_+-s_-}{s_-}}
		=& \frac 1{\pi\gamma^{\frac 32}} \frac{\sqrt{s_+-s_-}}{s_-}  \left(\frac{s_-}{s_+-s_-}\right)^{\frac 32} \Gamma \!\left( \frac 32 \right)
		= \frac 1{2\sqrt \pi} \frac{(s_-)^{\frac 12}}{(s_+-s_-)} \frac 1{\gamma^{\frac 32}}
	\end{align*}
	Similarly the contribution from the vicinity of $\lambda=s_+$ using the scaling $\lambda = s_+ - \frac u\gamma$ gives
	\begin{equation*}
		\frac 1{2\sqrt \pi} \frac{(s_+)^{\frac 12}}{(s_+-s_-)} \frac 1{\gamma^{\frac 32}}
	\end{equation*}
	and adding the two contributions gives
	\begin{equation*}
		\lim_{N\rightarrow\infty} \frac{\langle \mathcal N \rangle^{^+}}{N} \Biggr|_{\gamma >>1} \approx \frac 1{2\sqrt \pi} \frac{1}{\left(s_+^{\frac 12}-s_-^{\frac 12}\right)} \frac 1{\gamma^{\frac 32}}
	\end{equation*}
equivalent  \eqref{mainlimasy}.
\end{enumerate}

	\subsubsection{To section \ref{Edge Scaling}}.
 	Aiming to derive \eqref{av num edge sum} we start with \eqref{N+ delta after laplace n.1}and expand
in the exponential for small $\kappa$:
	\begin{align*}
		\mathcal L(\lambda)=&   \frac{\sqrt{\lambda^2+\kappa^2}}{\kappa} - \log\left[\frac{\kappa + \sqrt{\lambda^2+\kappa^2}}{\lambda}\right] - \lambda \frac{\sqrt{(\alpha-1)^2 + \kappa^2}}{(\alpha-1)\kappa} \\
		\approx& \frac \lambda\kappa \left(1+\frac{\kappa^2}{2\lambda^2} - \frac{\kappa^4}{8\lambda^4} \right)  - \log\left[1+\frac\kappa\lambda +\frac{\kappa^2}{2\lambda^2} - \frac{\kappa^4}{8\lambda^4} \right] - \frac\lambda\kappa \left(1 + \frac{\kappa^2}{2(\alpha-1)^2} + \frac{\kappa^4}{8(\alpha-1)^4} \right)+ \cdots \\
		=& \frac{\lambda}{\kappa}+\frac{\kappa}{2\lambda} - \frac{\kappa^3}{8\lambda^3}   - \left[\frac\kappa\lambda +\frac{\kappa^2}{2\lambda^2} -\frac 12\left( \frac\kappa\lambda +\frac{\kappa^2}{2\lambda^2} \right)^2 +\frac 13  \frac{\kappa^3}{\lambda^3} +o\left(\frac{\kappa^4}{\lambda^4}\right) \right] \\
		&-  \frac\lambda\kappa - \frac{\kappa\lambda}{2(\alpha-1)^2} - \frac{\kappa^3\lambda}{8(\alpha-1)^4} + \cdots \\
		=&-\frac{\kappa}{2}\left[  \frac{1}{\lambda}+ \frac{\lambda}{(\alpha-1)^2} \right] + \frac{\kappa^3}{8}\left[\frac{1}{3\lambda^3}      - \frac{\lambda}{(\alpha-1)^4}\right] + o\left(\frac{\kappa^4}{\lambda^4}\right)
	\end{align*}
	Next we recall $\kappa = (\alpha -1 ) \left( \delta + \frac{\delta^3}{6} + \cdots \right)$ and for $\delta << 1$ collect all terms up to third power in the exponent of \eqref{N+ delta after laplace n.1} is $\exp{-\frac N2 [ (\alpha+1)\delta + (\alpha-1) \mathcal L(\lambda) ] }$:
	\begin{align*}
		&(\alpha+1)\delta -\frac 12 (\alpha -1 ) \left( \delta + \frac{\delta^3}{6} \right) \left( \frac{\alpha-1}{\lambda} +\frac{\lambda}{\alpha-1}  \right) + \frac 18 (\alpha -1 )^3 \left( \delta + \frac{\delta^3}{6} \right)^3 \left( \frac{\alpha-1}{3\lambda^3} +\frac{\lambda}{(\alpha-1)^3}  \right)  \\
		&=\delta\left[(\alpha+1) - \frac{1}{2}\left( \frac{(\alpha-1)^2}{\lambda} +\lambda  \right)  \right]     + \frac {\delta^3 }{12} \left[ \frac \lambda2 +\frac{(\alpha -1 )^4}{2\lambda^3}   -  \frac{(\alpha-1)^2}{\lambda}    \right]
	\end{align*}
	To evaluate the contribution from the vicinity of the upper edge $s_+=(\sqrt{\alpha}-1)^2$
we substitute here $\lambda = s_+ + N^{-\frac 23}\left(\frac{4s_+^{2}}{(s_+-s_-)}\right)^{\frac 13} \xi$ as in
\eqref{edges} and reparametrize
\[
\delta=\frac{\omega s_{+}^{1/3}}{N^{1/3}} \left(\frac{s_+-s_-}{4}\right)^{-2/3}
\]
where both $\xi$ and $\omega_+$ are considered to be fixed as $N\to \infty$. After straightforward but cumbersome algebra we then find the corresponding contribution
	\begin{align*}
		\left\langle \mathcal N_\delta \right\rangle^{^+}_{s_+} \approx& 2 e^{-\frac 13 \omega_+^3} \int_{-\infty}^\infty e^{\omega_+ \xi} \rho_{edge} (\xi) d\xi  \\
	\end{align*}
	Similarly, the contribution around the lower edge $s_-$ is given by
	\begin{align*}
		\left\langle \mathcal N_\delta \right\rangle^{^+}_{s_-} \approx& 2 e^{-\frac 13 \omega_-^3} \int_{-\infty}^\infty e^{\omega_- \xi} \rho_{edge} (\xi) d\xi  \\
		&\mbox{where } \omega_- = N^{\frac 13} \delta \frac{[(s_+-s_-)/4]^{\frac 23}}{s_-^{\frac 13}}
	\end{align*}
The sum of the two contributions is equivalent to \eqref{av num edge sum}.

 For verifying \eqref{65} we write:
	\begin{align*}
		\left\langle \mathcal N_\omega \right\rangle^{^+} =& 2  \left[ \sqrt{\frac 2\pi} {x_+^{-\frac 32}} + \sqrt{\frac 2\pi} {x_-^{-\frac 32}} \right] = \frac { \left[  \sqrt{s_+} + \sqrt{s_-}\right]}{\omega^{\frac 32} \sqrt \pi } = \frac { 4 \left[  \sqrt{s_+} + \sqrt{s_-}\right]}{N^{\frac 12} \delta^{\frac 32} \left( s_+-s_- \right) \sqrt \pi } = \frac { 2 }{N^{\frac 12} \delta^{\frac 32}  \sqrt \pi }
	\end{align*}
which is  equivalent to \eqref{65}.

	\subsubsection{To  section \ref{Large deviations}}
\begin{enumerate}
\item	The expression for $\Phi ( \lambda)$ in \eqref{LD func} can be  written using
the large-deviation form for the density of Wishart eigenvalues outside the
Marchenko-Pastur support as given in  \cite{ForrLD}:
	\begin{align*}
		\Phi ( \lambda) =& (\alpha - 1) \left[ \frac{\sqrt{ \lambda^2+\kappa^2}}{\kappa} - \log \left( \frac{\kappa + \sqrt{ \lambda^2 + \kappa^2}}{ \lambda}\right) -  \lambda \frac{\sqrt{(\alpha - 1) + \kappa^2}}{(\alpha -1 ) \kappa}   \right]  \\
		&+ \left[ u_L ( \lambda) - 2 \log \left| \frac{u_L+ \lambda-1-\alpha}{2\sqrt \alpha} \right| + 2(\alpha -1)\log\left| \frac{u_L -  \lambda - \alpha + 1}{2\sqrt\alpha}  \right| \right] + (\alpha +1 )\delta \\
		\mbox{where } \kappa =& (\alpha-1)\sinh\delta = \frac{(\alpha-1)\sigma^2}{2\sqrt{1+\sigma^2}}, \quad \delta = \frac 12 \log(1+\sigma^2) \mbox{ and } u_L( \lambda) = - \sqrt{( \lambda - s_-)( \lambda - s_+)}.
	\end{align*}
 We can further simplify noticing that $\log  \lambda$  terms cancel and
	\begin{align*}
		& u_L <0, \, \lambda-1-\alpha < s_- - 1 -\alpha = -2 \sqrt\alpha <0\\
		&\Rightarrow |u_L + \lambda  - 1 - \alpha| = -u_L - \lambda + 1+\alpha\\
		& u_L <0, \, -\lambda - \alpha +1 < -u_L = \sqrt{( \lambda - s_-)( \lambda - s_+)} \\
		&\Rightarrow |u_L - \lambda   - \alpha+1| =  \lambda +\alpha - 1 - u_L
	\end{align*}
	
\item	To verify \eqref{49} we write the chain of identities, first averaging over the vector $\bm b$ and then over the Wishart matrices:
	\begin{align*}
		\left\langle\left\langle g(\lambda)\right\rangle_{\bm b}\right\rangle_{W} =& \left\langle\frac 1N \Tr{\frac{W}{(\lambda-W)^2}}\right\rangle  = - \left\langle \frac{\partial}{\partial \lambda} \left[ \frac 1N\Tr{\frac{W}{\lambda-W}} \right]\right\rangle  =- \left\langle \frac{\partial}{\partial \lambda} \left[ \frac 1N\Tr{\frac{\lambda}{\lambda-W}} \right]\right\rangle
\end{align*}
implying
\begin{align}
   \left\langle\left\langle g(\lambda)\right\rangle_{\bm b}\right\rangle_{W}  = - \frac{\partial}{\partial \lambda} \left[ \lambda G(\lambda) \right] \label{aveG}
	\end{align}
where $G(\lambda)$ is the mean trace of the resolvent of the Wishart matrix. Using the Marchenko-Pastur
density \eqref{MP} one can straightforwardly integrate to verify the well-known idenity valid for $\lambda < s_-$:
	\begin{equation}\label{Gres}
		G(\lambda): =  \frac 1N \left\langle \Tr{\frac{1}{\lambda - W}} \right\rangle \xrightarrow[N\rightarrow \infty]{} - \frac 2\lambda \frac{\sqrt{s_+ s_-} - \lambda - \sqrt{(\lambda - s_-)(\lambda-s_+)}}{(\sqrt{s_+}-\sqrt{s_-})^2}
	\end{equation}
	which when substituted to \eqref{aveG} gives
	\begin{align*}
		\left\langle\left\langle g(\lambda)\right\rangle_{\bm b}\right\rangle_{W} =& - \frac{2}{(\sqrt{s_+}-\sqrt{s_-})^2} \frac{\partial}{\partial \lambda} \left[   \lambda +
		\sqrt{\lambda^2 - \lambda (s_++s_-) + s_+s_-}
		\right] \\
		=& - \frac{1}{(\sqrt{s_+}-\sqrt{s_-})^2}  \left[
		\frac{2\sqrt{(s_+-\lambda)( s_- - \lambda )}  -[(s_+-\lambda) + (s_- -\lambda)]}{\sqrt{(s_+-\lambda)( s_- - \lambda )}} \right] \\
		=&  \frac{1}{(\sqrt{s_+}-\sqrt{s_-})^2}
		\frac{\left(\sqrt{s_+-\lambda} - \sqrt{s_- - \lambda }\right)^2}{\sqrt{(s_+-\lambda)( s_- - \lambda )}}
	\end{align*}
implying \eqref{49}.

\item 	to verify  \eqref{eq to 0} we write
	\begin{align*}
		\frac{dL_2}{d\lambda} =& \frac{\frac{s_++s_-}{2} - \lambda}{\sqrt{(\lambda - s_-)(\lambda - s_+)}} +  \frac{2}{\alpha +1 -\lambda +\sqrt{(\lambda - s_-)(\lambda - s_+)}} \left(1+\frac{\frac{s_++s_-}{2} - \lambda}{\sqrt{(\lambda - s_-)(\lambda - s_+)}}  \right)\\
		&+ \frac{2(\alpha - 1)}{\alpha -1 +\lambda +\sqrt{(\lambda - s_-)(\lambda - s_+)}} \left(1-\frac{\frac{s_++s_-}{2} - \lambda}{\sqrt{(\lambda - s_-)(\lambda - s_+)}}  \right)\\
		=&\frac{\frac{s_++s_-}{2} - \lambda}{\sqrt{(\lambda - s_-)(\lambda - s_+)}} \left[ 1+\frac{2}{\alpha +1 -\lambda +\sqrt{(\lambda - s_-)(\lambda - s_+)}} - \frac{2(\alpha-1)}{\alpha -1 +\lambda +\sqrt{(\lambda - s_-)(\lambda - s_+)}}  \right] \\ &+ \frac{2}{\alpha +1 -\lambda +\sqrt{(\lambda - s_-)(\lambda - s_+)}} + \frac{2(\alpha-1)}{\alpha -1 +\lambda +\sqrt{(\lambda - s_-)(\lambda - s_+)}}
	\end{align*}

We evaluate this using the expression for $\lambda = \lambda_\ast=\alpha+1-\sqrt\alpha \left( \sqrt{1+\sigma^2} + \frac{1}{\sqrt{1+\sigma^2}} \right)$. First we find that 	
	\begin{align*}
		\sqrt{(\lambda_\ast - s_-)(\lambda_\ast - s_+)} = \sqrt \alpha \frac{\sigma^2}{\sqrt{1+\sigma^2}}
\end{align*}
which gives after straightforward algebra
\begin{align*}
		& \alpha - 1 + \lambda_\ast + \sqrt{(\lambda - s_-)(\lambda - s_+)} = \frac{2\sqrt{\alpha} \left(\sqrt{\alpha (1+\sigma^2)}-1\right)}{\sqrt{1+\sigma^2}},\\
		& \alpha + 1 - \lambda_\ast + \sqrt{(\lambda - s_-)(\lambda - s_+)} = 2\sqrt{\alpha}\sqrt{1+\sigma^2}, \quad   \frac{\frac{s_++s_-}{2}}{\sqrt{(\lambda - s_-)(\lambda - s_+)}} = \frac{2+\sigma^2}{\sigma^2}.
	\end{align*}
and using the above
	\begin{align*}
		\frac{dL_2}{d\lambda}\Big|_{\lambda_\ast} =& \frac{2+\sigma^2}{\sigma^2} \left[  1+ \frac{1}{\sqrt{\alpha (1+\sigma^2)}} - \frac{(\alpha-1)\sqrt{1+\sigma^2}}{\sqrt{\alpha} \left(\sqrt{\alpha (1+\sigma^2)}-1\right)} \right] \\
		&+\frac{1}{\sqrt{\alpha (1+\sigma^2)}} + \frac{(\alpha-1)\sqrt{1+\sigma^2}}{\sqrt{\alpha} \left(\sqrt{\alpha (1+\sigma^2)}-1\right)} \\
		=& \frac{2+\sigma^2}{\sigma^2} - \frac{2(\alpha-1)\sqrt{1+\sigma^2}}{\sigma^2\sqrt{\alpha} \left(\sqrt{\alpha (1+\sigma^2)}-1\right)} +\frac{2\sqrt{1+\sigma^2}}{\sigma^2\sqrt{\alpha}}
	\end{align*}
	Further, we evaluate for $L_1$
	\begin{align*}
		\frac{dL_1}{d\lambda} =& (\alpha - 1) \left[ \frac{\lambda}{\kappa \sqrt{\kappa^2 + \lambda^2}} - \frac{1}{\kappa+ \sqrt{\kappa^2 + \lambda^2}}\frac{\lambda}{ \sqrt{\kappa^2 + \lambda^2}} - \frac{\sqrt{ (\alpha-1)^2 + \kappa^2 }  }{(\alpha -1)\kappa} \right]  \\
		=& (\alpha-1) \frac{\lambda}{\sqrt{\kappa^2 + \lambda^2}} \left[\frac 1\kappa - \frac 1{\kappa + \sqrt{\kappa^2 + \lambda^2}}\right] - \frac{\sqrt{ (\alpha-1)^2 + \kappa^2 }  }{\kappa}  \\
		=& \frac{(\alpha-1)\lambda}{\kappa \left( \kappa + \sqrt{\kappa^2 + \lambda^2}\right)} - \frac{\sqrt{ (\alpha-1)^2 + \kappa^2 }  }{\kappa} = \frac{(\alpha-1)\lambda\left( \kappa + \sqrt{\kappa^2 + \lambda^2}\right)}{\kappa \lambda^2 } - \frac{\sqrt{ (\alpha-1)^2 + \kappa^2 }  }{\kappa}  \\
		=& \frac{(\alpha-1)\sqrt{\kappa^2 + \lambda^2}}{\kappa \lambda } - \frac{\alpha -1}{\lambda} - \frac{\sigma^2+2}{\sigma^2}
	\end{align*}
	Therefore,
	\begin{align*}
		\left( \frac{dL_1}{d\lambda} +\frac{dL_2}{d\lambda} \right)\Biggr|_{\lambda_\ast} =& \frac{2\sqrt{1+\sigma^2}}{\sigma^2 \sqrt{\alpha}} - \frac{2(\alpha-1)\sqrt{1+\sigma^2}}{\sigma^2 \sqrt{\alpha} \left( \sqrt{\alpha (1+\sigma^2)}-1 \right) }
		+\frac{(\alpha-1)\sqrt{\kappa^2+\lambda^2_\ast}}{\kappa \lambda_\ast} - \frac{\alpha -1}{\lambda_\ast}  \\
		=& \frac{2\sqrt{1+\sigma^2}}{\sigma^2 \sqrt{\alpha}} \left[ 1-\frac{\alpha -1}{\sqrt{\alpha (1+\sigma^2)} -1} \right] + \frac{(\alpha-1)\sqrt{\kappa^2+\lambda^2_\ast}}{\kappa \lambda_\ast} - \frac{\alpha - 1}{\lambda_\ast}  \\
		=&  \frac{2\sqrt{1+\sigma^2}}{\sigma^2 \sqrt{\alpha}} \frac{\sqrt{\alpha (1+\sigma^2)}-\alpha}{\sqrt{\alpha (1+\sigma^2)} -1} + \frac{(\alpha-1)\sqrt{\kappa^2+\lambda^2_\ast}}{\kappa \lambda_\ast} - \frac{\alpha - 1}{\lambda_\ast}  \\
		=& - \frac{2\sqrt{1+\sigma^2}\left(\sqrt{ 1+\sigma^2}-\sqrt\alpha\right)^2}{\sigma^2 \left(\sqrt{\alpha (1+\sigma^2)} -1\right)\left(\sqrt \alpha -\sqrt{ 1+\sigma^2}\right)}  + \frac{(\alpha-1)}{\lambda_\ast}\left[\frac{\sqrt{\kappa^2+\lambda^2_\ast}}{\kappa } -1  \right] \\
		=& - \frac{2 \left(\sqrt{ 1+\sigma^2}-\sqrt\alpha\right)^2}{\lambda_\ast}  + \frac{(\alpha-1)}{\lambda_\ast}\left[\frac{\sqrt{\kappa^2+\lambda^2_\ast}}{\kappa } -1  \right]
	\end{align*}
which is equivalent to \eqref{eq to 0}
\item	For verifying that \eqref{eq to 0} vanishes we first
	assume for simplicity $\lambda_\ast \neq 0$, i.e. $\sigma^2 \neq \alpha - 1$ but actually one will see this condition is immaterial. Then the problem amounts to verifying the identity:
	\begin{equation*}
		\frac{2}{\sigma^2} \left(\sqrt{ 1+\sigma^2}-\sqrt\alpha\right)^2 =(\alpha - 1)\left[\frac{\sqrt{\kappa^2+\lambda^2_\ast}}{\kappa } -1  \right]
	\end{equation*}
	for which we rewrite  equivalently as
	\begin{align*}
		 \frac{2}{\sigma^2} \frac{\left(\sqrt{ 1+\sigma^2}-\sqrt\alpha\right)^2}{\alpha-1} +1 = \sqrt{1+\frac{\lambda^2_\ast}{\kappa^2}}
\end{align*}
We can further square both parts as they are positive and substitute the definition of $\kappa$:
\begin{align*}
		& \left[ \frac{2}{\sigma^2} \frac{\left(\sqrt{ 1+\sigma^2}-\sqrt\alpha\right)^2}{\alpha-1} +1 \right]^2 -1 = \frac{\lambda^2_\ast}{\kappa^2} = \lambda^2_\ast \frac{4 (1+\sigma^2)}{(\alpha - 1)^2\sigma^4},\\
		\Leftrightarrow\,\, & \left[ \frac{2}{\sigma^2} \frac{\left(\sqrt{ 1+\sigma^2}-\sqrt\alpha\right)^2}{\alpha-1} +2 \right] \frac{2}{\sigma^2} \frac{\left(\sqrt{ 1+\sigma^2}-\sqrt\alpha\right)^2}{\alpha-1} = \lambda^2_\ast \frac{4 (1+\sigma^2)}{(\alpha - 1)^2\sigma^4},
	\end{align*}
and upon further rearranging
	\begin{align*}
	\Leftrightarrow \,\,	& \left[ \left(\sqrt{ 1+\sigma^2}-\sqrt\alpha\right)^2 + \sigma^2(\alpha-1)  \right] \frac{\left(\sqrt{ 1+\sigma^2}-\sqrt\alpha\right)^2}{1+ \sigma^2} = \lambda^2_\ast, \\
	\Leftrightarrow\,\,	&  \frac{\left(\sqrt{\alpha (1+\sigma^2)} - 1  \right)^2 \left(\sqrt{ 1+\sigma^2}-\sqrt\alpha\right)^2}{1+ \sigma^2} = \lambda^2_\ast
	\end{align*}
	the last equation being indeed true.  Therefore $\lambda_\ast$ is a stationary point (actually, the minimum) of the large deviations function $\Phi (\lambda)$.

\end{enumerate}

\subsection{Averaging of the replicated partition function for a general least-square random cost}\label{partion_fun_evaluation}
In this Appendix we show how to evaluate the replicated moments of the finite-temperature partition function induced by a general least-square random cost landscape of the form
\begin{equation}\label{cost_gen_poly}
		H (\bm x) = \frac{1}{2} \sum\limits_{k=1}^M \left[V_k(\bm x)\right]^2 (\bm{x}).
	\end{equation}
restricted to the sphere ${\bm x}^2=N$, where $V_k(\bm{x})$ are chosen to be gaussian-distributed functions of the vector $x$ with expectations $\mathbb E\{V_k\}=0$ and covariance
\begin{equation}\label{cov_gen}
\mathbb E \{V_k(\bm x_a) V_l (\bm x_b)\} = \delta_{kl} f\! \left(\frac{\bm x_a \cdot \bm x_b}{N} \right)
\end{equation}
The particular case $V_k(\bm x)=(\bm a_k,\bm x)-b_k$ treated in this paper corresponds to the choice $f(u)=\sigma^2+u$, but derivation can be done for any covariance of the form \eqref{cov_gen}
and essentially follows the method of  \cite{Fyo2019}.

	Starting with the partition function $Z$ as defined in the Introduction and repeated below
	\begin{equation}\label{method Z_Appendix1}
		Z = \int\limits_{{\bm x}^2=N} d \bm x e^{-\frac 12 \beta  \sum\limits_{k=1}^M V_k^2 (\bm x)}
	\end{equation}
	we aim to arrive to a convenient integral representation for the averaged partition function of $n$ copies of the same system $\langle Z^n \rangle$.	To facilitate the averaging we introduce new auxiliary variables of integration and employ the standard gaussian integral identity (sometimes called in physics literature the Hubbard-Stratonovich identity)	 applying it $M$ times for every index $k=1,\ldots,M$. Proceeding in this way the partition function $Z$ now can be expressed in terms of the integration over a vector ${\bm u = (u_1, \dots , u_M)^T}$ as
	\begin{equation}
		Z = \int\limits_{\mathbb R^M} \frac{d \bm u}{ (2\pi)^{M/2}} e^{-\frac 12 (\bm u \cdot  \bm u)} \int\limits_{{\bm x}^2=N} d \bm x e^{-i \sqrt{\beta} \sum\limits_{k=1}^M u_k V_k (\bm x)}
	\end{equation}
	where $(\bm u \cdot \bm v)$ stands for the scalar product.	

 Now we	take products of identical copies of $Z$ (which we number with the index $a=1,2,\ldots n$) and aim at evaluating
	\begin{equation}
		\langle Z^n \rangle =  \int\limits_{\mathbb R^{nM}} \prod_{a=1}^n \frac{d \bm u_a}{ (2\pi)^{M/2}} e^{-\frac{1}{2} \sum\limits_{a=1}^n (\bm u_a \cdot \bm u_a)}  \idotsint\limits_{D_N} \prod\limits_{a=1}^n d \bm x_a \prod\limits_{k=1}^M\left\langle e^{-i \sqrt{\beta}  \sum\limits_{a=1}^n [\bm u_a]_k V_k (\bm x_a)} \right\rangle_V
	\end{equation}
	where rectangular brackets $[\bm u_a]_k$ stand for the $k$-th entry of the vector $\bm u_a$. Above we used that
for different $k$ the functions $V_k(x)$ are independent, hence the corresponding average of the product factorizes in the product of averages. The domain of integration is the union of $n$ spheres $D_N = \{  \, {\bm x_a}^2 = N, \,  \forall a \}$.

	 To perform the average over $V_k$  we use that the combination
$z=-i \sqrt{\beta}  \sum\limits_{a=1}^n [\bm u_a]_k V_k (\bm x_a)$ is Gaussian with mean zero and the variance
\[
\mathbb E \{z^2\}=-\beta\sum\limits_{a,b=1}^n [\bm u_a]_k[\bm u_b]_k E \{V_k (\bm x_a)V_k (\bm x_b)\}
\]
which upon using the covariances \eqref{cov_gen} and taking the product over $k=1,\ldots,M$
yields
	\begin{align}
		\prod\limits_{k=1}^M \langle e^{-i \sqrt{\beta} \sum\limits_{a=1}^n [\bm u_a]_k V_k (\bm x_a)} \rangle_V
		&= \prod\limits_{k=1}^M \exp\left[-\frac 12 \beta \sum_{a,b=1}^n [\bm u_a]_k [\bm u_b]_k f\! \left(\frac{\bm x_a \cdot \bm x_b}{N} \right) \right]\\
		&= \exp\left[-\frac 12 \beta (\bm u_a \cdot \bm u_b) f\! \left(\frac{\bm x_a \cdot \bm x_b}{N} \right) \right]
	\end{align}
Substituting this back to the integral $\langle Z^n \rangle$ we notice that the integral over vectors $\bm u_a$  is simply a multivariate  $n$-dimensional gaussian integration, taken $M$ times:
	\begin{align}
		\langle Z^n \rangle
		&= \idotsint\limits_{D_N} \prod\limits_{a=1}^n d \bm x_a \int\limits_{\mathbb R^{nM}} e^{-\frac 12 (\bm u_1, \dots , \bm u_n) \left[I_n + \beta f \left(\frac{ \bm x_a \cdot \bm x_b }{N} \right) \right] (\bm u_1, \dots , \bm u_n)^T } d \bm u_1 \dots d{\bm u}_n\notag \\
		&\propto \idotsint \limits_{D_N} \prod\limits_{a=1}^n d \bm x_a \left(\det \left[I_n + \beta f \left(\frac{ \bm x_a \cdot \bm x_b}{N} \right) \right] \right)^{-\frac M2} \notag\\
		&\propto \int\limits_{D_N^{(Q)}} d Q (\det Q)^{\frac{N-n-1}{2}} \left( \det \left[I_n + \beta \hat f (Q ) \right] \right)^{-\frac M2} \label{<Z^n> pre}
	\end{align}
	Here we omitted the exact proportionality constants (which are known but redundant for our goals). The change of variables in the last line  from the set of vectors ${\bm x_a}$ to the positive (semi)definite matrix $Q$ with entries defined as $q_{ab}=\frac 1N (\bm x_a \cdot \bm x_b)$ follows the idea of the paper \cite{Fyo2002}. The details of this transformations, including evaluation of the involved Jacobian determinant factor appearing in the above are explained in detail in ~\cite{FyoPhysA2010}, eq.(47). The hat over $\hat f (Q)$ serves as a reminder that this is an $n \times n$ matrix with entries $f_{ab}:=f(q_{ab})$. Finally, we rewrite the integral $\langle  Z^n \rangle$ in the form convenient for approximating it in the limit $N\gg 1$ by Laplace's method:
	\begin{align*}
		\langle Z^n \rangle &\propto \int\limits_{D_N^{(Q)}} d Q \, (\det Q)^{-\frac{n+1}{2}} e^{  -\frac N2 \Phi_n (Q) } ,\\
		&\Phi_n(Q) =\alpha \log \det \{ I_n + \beta \hat f (Q )  \} - \log \det Q
	\end{align*}
	which for the particular choice $f(u)=\sigma^2+u$ is equivalent to \eqref{special case}  of the main text.

\end{document}